\documentclass[3p,11pt]{elsarticle}

\usepackage{hyperref}
\usepackage{amssymb,amsmath}

\usepackage{amsthm}

\usepackage{bbm}
\usepackage{caption}
\usepackage{subcaption}
\usepackage{xcolor}
\usepackage{setspace}
\usepackage{graphicx}

\usepackage{rotating}
\usepackage{hvfloat}

\usepackage{tikz} 
\usetikzlibrary{shapes.geometric, arrows}   
\usetikzlibrary{shadows,arrows.meta,positioning,backgrounds,fit, arrows}

\tikzset{%
	materia/.style={draw, fill=blue!20, text width=8.0em, text centered, minimum height=1.5em,drop shadow},
	materia1/.style={draw, fill=blue!20, text width=8.0em, text centered, minimum height=1.5em,drop shadow},
	etape/.style={materia, text width= 10em, minimum width=10em, minimum height=3em, rounded corners, drop shadow},
	texto/.style={above, text width=6em, text centered},
	linepart/.style={draw, thick, color=black!50, -LaTeX, dashed},
	line/.style={draw, thick, color=black!50, -LaTeX},
	ur/.style={draw, text centered, minimum height=0.01em},
	back group/.style={fill=yellow!20,rounded corners, draw=black!50, dashed, inner xsep=15pt, inner ysep=10pt},
	arrow/.style = {-{Stealth[scale=1.2]}, rounded corners, thick},
	 decision/.style = {materia1, rectangle, text width= 12em, minimum width=12em, minimum height=4em},
edge from parent/.style= {draw, arrow},
}

\newcommand{\etape}[2]{node (p#1) [etape] {#2}}

\newcommand{\decision}[2]{node (p#1) [decision] {#2}}

\usepackage{subcaption}
\usepackage{bm}

\usepackage{booktabs}

\makeatletter
\def\ps@pprintTitle{%
\let\@oddhead\@empty
\let\@evenhead\@empty
\def\@oddfoot{}%
\let\@evenfoot\@oddfoot}
\makeatother

\begin{document}

\begin{frontmatter}

\title{Investigating Short-Term Dynamics in Green Bond Markets}


\author{Lorenzo Mercuri}
\address{Department of Economics Management and Quantitative Methods, University of Milan, Milan, Italy \\ email: \texttt{lorenzo.mercuri@unimi.it}}

\medskip

\author{Andrea Perchiazzo}
\address{Faculty of Economics and Social Sciences and Solvay Business School, Vrije Universiteit Brussel, Brussels, Belgium\\ email: \texttt{andrea.perchiazzo@vub.be}}

\medskip

\author{Edit Rroji}
\address{Department of Statistics and Quantitative Methods, University of Milano-Bicocca, Milan, Italy \\ email: \texttt{edit.rroji@unimib.it}}

%
%

\begin{abstract}
The paper investigates the effect of the label green in bond markets from the lens of the trading activity.  The idea is that jumps in the dynamics of returns have a specific memory nature that can be well represented through a self-exciting process. Specifically, using Hawkes processes where the intensity is described through a continuous time moving average model, we study the high-frequency  dynamics of bond prices. We also introduce a bivariate extension of the model that deals with the cross-effect of upward and downward price movements. 
Empirical results suggest that differences emerge if we consider periods with relevant interest rate announcements, especially in the case of an issuer operating in the energy market. 

\end{abstract}


\end{frontmatter}


\section*{Introduction}

In recent years the pressure of solutions for the environment (e.g., transition to a lower-carbon economy, energy efficiency, and construction of renewable source plants) has solicited the need of alternative funding. A clear example is the creation of a financial instrument called green bond, which is used exclusively to raise money for financing environmental (labeled as ``green'') projects. Following the principles of the International Capital Markets Association (ICMA) some categories of these investments funded through green bonds are destined to renewable energy, green buildings, public transport, energy efficiency, water, and waste management. 

A topic of interest in the financial literature is the ``greenium'' (or ``green'' premium) effect, which refers to the fact that bondholders are willing to pay a higher price or accept a lower yield in order to invest in green securities compared to conventional securities with similar characteristics (e.g., maturity, coupon rate, and credit profile of the issuer). 

Several papers have tried to identify the main factors that drive the ``greenium'' effect with particular emphasis on the issuer (that is, differentiating corporate, municipal, and sovereign bonds). Hereafter, we list some findings. \cite{hachenberg2018} do not find significant differences between the yield of green and conventional bonds with the same issuer; whereas, for example, \cite{bachelet2019} find that the type of bond issuer (i.e., institutional versus private) is relevant in order to detect the ``greenium'' effect. In addition to that, \cite{baker2018} claim that US corporate and municipal green-labeled bonds are issued above par (that is, at premium) with respect to non-green-labeled bonds, while \cite{reboredo2018} and \cite{pham2020} show that green bonds provide diversification benefits to portfolios. Moreover, \cite{martiradonna2023beneficial} suggest that portfolios with green bonds are less risky through time for different allocation strategies. \cite{sohag2023green} show that the co-movements between green-labeled and non-green-labeled markets are more pronounced in the short-term horizon.

Previous literature is mainly based on panel regressions that consider weekly or daily data (see for example \cite{pietsch2022pricing}) in which contract-specific or general macroeconomic variables are included as explanatory variables. Green bonds started to be issued in European Union only in the last decade when interest rates were low or negative for a long period of time and, consequently, the existing literature refers to markets with those conditions. In this paper, we propose a different viewpoint as we consider the intraday dynamics of bond returns where jumps are potentially explained through some market announcements.
 
Focusing on the high frequency framework, classical bonds have been widely investigated in the literature (see, for instance, \cite{biais2019microstructure} and \cite{bessembinder2020survey}) whereas, at the best of our knowledge, a study on green bonds has not been taken under consideration yet. In this paper, we analyze the effect of the label green by looking at returns computed on tick data  and
we propose a procedure for the investigation of the impact news in the market may have on the time series of bond prices. In particular, we use the Lee and Mykland's (LM) test (see \cite{lee2008jumps} for further details)  for the identification of jumps. The framework presented in the paper can be generalized by considering alternative tests for jump detection, see for example \cite{dumitru2012identifying} for a list of non-parametric tests.

Jumps are important in finance from different points of view. For example \cite{brignone2023} show how to price Asian options in presence of jump clusters while \cite{li2022oil} show that jumps significantly affect future realized volatility and used them for oil futures volatility prediction.  Our analysis is based on a counting process, proposed in \cite{mercuri2022hawkes}, where the conditional intensity is modelled with the continuous version of the ARMA process called a CARMA(p,q)-Hawkes. Such a model implies a self-exciting behaviour with a non monotone autocorrelation function for jumps in non-overlapping intervals of the same length. We expect some particular market conditions to generate different sequences of jumps for pairs of brown and green bonds with similar features in terms of credit quality, coupon rate, duration, and currency. As a set of jumps defines a point process, then by distinguishing between positive and negative jumps it is possible to consider a bivariate point process. To this aim, we introduce the bivariate version of the CARMA(p,q)-Hawkes model
and present an estimation procedure.
In our empirical analysis, we consider two periods each of length one month. In the first period (from August 19 to September 19, 2022) the market dealt with an announcement of the European Central Bank (ECB) on the interest rate raise while in the second period (from November 14 to December 13, 2022) the market had no announcements from ECB but it experienced a particularly nervous moment due to the  uncertainty in gas price in Europe. Issuers are two companies that operate in the European market. The first is the financial institution Cr\'edit Agricole SA  while the second issuer, which is a corporate company operating in the energy sector issuing both green and brown bonds, is Engie SA.\footnote{Engie SA is the largest non sovereign green bond issuer in France.}

This paper attempts to shed light on differences that emerge in terms of market activity and liquidity for different bond redemption types, market sector of the issuer and currency. 


In our analyses, bid prices display more jumps for the series considered. We also observe that higher order CARMA(p,q)-Hawkes  seems to be useful for the series of jumps extracted from the returns of the bonds issued by the corporate company operating in the energy sector. 

The paper is organized as follows. Section~\ref{models} introduces the bivariate CARMA(p,q)-Hawkes model. Section~\ref{estimation} presents the likelihood function of the bivariate CARMA(p,q)-Hawkes model, the different frameworks under investigation and the description of the estimation procedure. Section \ref{empirical} shows the empirical results. Section \ref{concl} concludes the paper.

\section{Self-exciting processes}\label{models}

We briefly review classical counting processes and discuss the main features of a CARMA(p,q)-Hawkes model\footnote{As highlighted in \cite{mercuri2022hawkes}, an exponential Hawkes process shows a strictly decreasing behaviour of the autocorrelation function. In order to remove such a monotonicity constraint,  it is possible to consider a Hawkes process where the intensity follows a CARMA process that can provide several shapes of the autocorrelation function. The latter is an extension of the classical Hawkes model.} (see \cite{mercuri2022hawkes} for a complete overview) that is an extension of the Hawkes type processes where the effect of past realizations is not monotonically decreasing, see \cite{bacry2015hawkes} for an overview of Hawkes processes in finance. In order to investigate the cross effect of positive and negative jumps in the series of bond returns we introduce the bivariate CARMA(p,q)-Hawkes model. 

\subsection{Hawkes-type processes}\label{hawkes_proc}

A point process, which is used to describe the dynamics of observed event times, is a collection of realizations $\{t_i\}_{i=1}^{\infty}$ where $t_0:=0$ and $t_i\geq 0$ for $i=1,2, \ldots$ of the non-decreasing and non-negative process $\left\{T_{i}\right\}_{i \geq 1 }$ (i.e., the time arrival process). 
The counting process $N_t$, representing the number of events up to time $t$ and obtained from the time arrival process, is defined as
\begin{equation}
N_{t} :=\sum_{i \geq 1} \mathbf{1}_{\{T_i\leq t\}}
\label{genCountingProcess}
\end{equation}  
for $t \geq 0$ with associated filtration $(\mathcal{F}_t)_{t\geq 0}$, which contains the information of $N_t$. In practice, if we look at a series of returns the point process refers to the sequence of time events while the counting process to the number of jumps occur up to a fixed time instant $t$.   
The conditional intensity $\lambda_t$ for the counting process $N_t$ is defined as
\begin{equation*}
\lambda_t=\lim_{\Delta\rightarrow 0^+}\frac{\mathsf{Pr}[N_{t+\Delta}-N_t=1|\mathcal{F}_{t}]}{\Delta},
\end{equation*}
and straightforwardly follows that the counting process satisfies
\begin{equation*}
\mathsf{Pr}\left[N_{t+\Delta}-N_t= \eta \left|\mathcal{F}_t\right.\right] =
\begin{cases}
    1-\lambda_t \Delta + o\left(\Delta\right)       &  \quad \text{if } \eta = 0 \\
    \lambda_t \Delta + o\left(\Delta\right)          & \quad \text{if } \eta = 1 \\
		o\left(\Delta\right)                             & \quad \text{if } \eta > 1.
\end{cases}
\end{equation*}
The conditional intensity $\lambda_t$ of a general self-exciting process has the following form
\begin{equation}
\lambda_t = \mu + \int_{0}^{t}h\left(t-s\right)\mbox{d}N_s
\label{CondIntensity}
\end{equation}
with baseline intensity parameter $\mu > 0$ and kernel function $h\left(t\right):\left[0,+\infty\right)\rightarrow \left[0,+\infty\right)$, representing the contribution to the intensity at time $t$ made by an event occurred at a previous time $T_i < t$. Note that the condition of stationarity  is given by $\int_0^{+\infty}h\left(t\right)\mbox{d}t<1$.
The case of the exponential kernel, i.e. $h\left(t\right)=\alpha e^{-\beta t}$ with $\alpha,\beta\geq 0$, defines the classical Hawkes model where the long-run autocorrelation function of the number of jumps over intervals of length $\tau$ separated by a time lag of $\delta$ has the following characteristics: i) is strictly monotonic decreasing function of $\delta$ (see \cite{da2014hawkes} for further details), ii) is always positive for $\alpha<\beta$ (stationarity condition), and iii) is exponentially decaying with lag $\delta$.  The use of the continuous autoregressive moving average models for the conditional dynamics of $\lambda_t$ generalizes the classical Hawkes with the advantage of reproducing different shapes of the autocorrelation function. 

\subsection{CARMA(p,q)-Hawkes process}\label{Hawkes}

A \((p+1)\) vector process \([X_{1,t},\ldots,X_{1,p},N_{t}]^{\top}\) is classified as a CARMA(p,q)-Hawkes process if the conditional intensity \(\lambda_t\) of the counting process \(N_t\) takes the form of a CARMA(p,q) process driven by \(N_t\), characterized as follows
\begin{equation}
\lambda_t = \mu+ \mathbf{b}^\top X_t.
\label{CondIntCH}
\end{equation}
\(\mu\) is the positive baseline intensity, and \(\mathbf{b}=[b_0,b_1,\ldots,b_{p-1}]^\top\) with \(b_{q+1}=\ldots=b_{p-1}=0\) contains the moving average parameters. The vector \(X_{t}=[X_{1,t},\ldots,X_{1,p}]^\top\) satisfies the subsequent linear stochastic differential equation
\begin{equation}
\mbox{d}X_t = \mathbf{A}X_{t-}\mbox{d}t+\mathbf{e}\mbox{d}N_t \text{ with } X_0=\mathbf{0},
\label{eq:CarLambda}
\end{equation}
where \(\mathbf{e}=[0,0,\ldots,1]^\top\) and the matrix \(\mathbf{A}\) is defined as
\begin{equation}
\mathbf{A}=\begin{bmatrix}
0 & 1 & 0 & \ldots & 0\\
0 & 0 & 1 & \ldots & 0\\
\vdots & \vdots & \vdots & \ddots & \vdots\\
0 & 0 & 0 & \ldots & 1\\
-a_p & -a_{p-1} & -a_{p-2} & \ldots & -a_1\\
\end{bmatrix}_{p\times p},
\label{Acomp}
\end{equation}
where $\left(a_1, \ldots, a_p\right)$ are the autoregressive parameters.\newline
The stochastic differential equation in \eqref{eq:CarLambda} admits an analytical solution given the initial condition, which reads: 
\begin{equation}
X_t=\int_{0}^{t} e^{\mathbf{A}\left(t-s\right)}\mathbf{e} \mbox{d}N_s.
\label{solXt}
\end{equation}

We now provide an overview of the likelihood function for the CARMA(p,q)-Hawkes process (for detailed insights, refer to \cite[Section 4.2]{mercuri2022hawkes}) that we will use in the empirical analysis. Given observed time arrivals \(T_1, \dots, T_n\), the likelihood function is expressed as
\begin{equation*}
	\mathcal{L}\left(\theta\right) = -\mu T_n - \mathbf{b}^\top \mathbf{A}^{-1} S \left(n\right)\mathbf{e} + n \mathbf{b}^\top \mathbf{A}^{-1} \mathbf{e} + \sum_{i = 1}^{n}\ln\left(\lambda_{T_i}\right),
\end{equation*}
where \(S \left(n\right)  = \sum_{i = 1}^{n} e^{\mathbf{A}\left(T_n - T_i\right)}\).\newline
The stationarity conditions for the CARMA(p,q)-Hawkes, ensuring a non-negative kernel, and establishing the diagonalizability of matrix \(\mathbf{A}\) are discussed in \cite{mercuri2022hawkes}. The same paper also offers a simulation algorithm and an alternative estimation procedure based on the autocorrelation function of the number of jumps observed in two non-overlapping intervals of equal length.

\subsection{Bivariate CARMA(p,q)-Hawkes process}\label{CARMAHawkes}
Let $N_{t}^+$ and $N_{t}^-$ denote respectively the number of upward and downward jumps in the time interval $[0, t]$. The bivariate version of a CARMA(p,q)-Hawkes process imposes a cross-dependence between the upward and downward jump intensities respectively named $\lambda_{t, 1}$ and $\lambda_{t, 2}$. As the number of parameters increases in the bivariate case, we introduce the following quantities for sake of clarity in the model description: 
\begin{enumerate}[i)]
	\item let $\mathbf{p} := \left[p_1, p_2\right]$ be the dimension of the autoregressive parameters where $p_1$ and $p_2$ are respectively the dimensions of the state processes $X_{t,1}$ and $X_{t,2}$;
	\item let $\mathbf{q} := \left[q_1, q_{1,2}, q_{2,1}, q_{2} \right]$ be the dimension of the moving average parameters;
	\item let $\mathbf{b}_{1,1}:= \left[ b_{1,1}^{(0)}, \dots, b_{1,1}^{(p_1 - 1)} \right]^\intercal$ with $b_{1, 1}^{(q_1+ 1)} = \dots = b_{1,1}^{(p_1 - 1)} = 0$ be a column vector of dimension $p_{1}\times 1$;
	\item let $\mathbf{b}_{1,2}:= \left[ b_{1,2}^{(0)}, \dots, b_{1,2}^{(p_2 - 1)} \right]^\intercal$ with $b_{1, 2}^{(q_{1,2} + 1)} = \dots = b_{1,2}^{(p_2 - 1)} = 0$ be a column vector of dimension $p_{2}\times 1$;
	\item let $\mathbf{b}_{2,1}:= \left[ b_{2,1}^{(0)}, \dots, b_{2,1}^{(p_1 - 1)} \right]^\intercal$ with $b_{2, 1}^{(q_{2,1} + 1)} = \dots = b_{2,1}^{(p_1 - 1)} = 0$ be a column vector of dimension $p_{1} \times 1$;
	\item let $\mathbf{b}_{2,2}:= \left[ b_{2,2}^{(0)}, \dots, b_{2,2}^{(p_2 - 1)} \right]^\intercal$ with $b_{2, 2}^{(q_2+ 1)} = \dots = b_{2,2}^{(p_2 - 1)} = 0$ be a column vector of dimension $p_{2}\times 1$.
\end{enumerate}
The intensities of $N_{t}^+$ and $N_{t}^-$, respectively $\lambda_{t, 1}$ and $\lambda_{t, 2}$, are modelled as
\begin{equation}
	\begin{split}
		\lambda_{t, 1} &= \mu_1 + \mathbf{b}_{1,1}^\intercal  X_{t,1} + \mathbf{b}_{1,2}^\intercal  X_{t,2} \ \\
		\lambda_{t, 2} &= \mu_2 + \mathbf{b}_{2,1}^\intercal  X_{t,1} + \mathbf{b}_{2,2}^\intercal  X_{t,2}
	\end{split}
\end{equation}
where
\begin{equation}
	\begin{split}
		\mathrm{d}X_{t,1} &=  \mathbf{A}_{1} X_{t,1} \mathrm{d}t + \mathbf{e}_1 \mathrm{d}N_{t}^+ \\
		\mathrm{d}X_{t,2} &=  \mathbf{A}_{2} X_{t,2} \mathrm{d}t + \mathbf{e}_2 \mathrm{d}N_{t}^-.
	\end{split}
\end{equation}
It is worth to note that $\mathbf{A}_{1}$, $\mathbf{A}_{2}$, $\mathbf{e}_1$, and $\mathbf{e}_2$ have the same structure of the corresponding $\mathbf{A}$ and $\mathbf{e}$ denoted in Section~\ref{Hawkes}.

\noindent Then $\mathbf{N}_t:= \left[N_{t}^+ \hspace{0.4em} N_{t}^- \right]^{\top}$ is a bivariate self-exciting counting process with intensity process $\bm{\lambda}_t = \left[\lambda_{t, 1} \hspace{0.4em} \lambda_{t, 2} \right]^{\top}$. It is possible to simplify the notation using the following quantities
\begin{equation*}
	\bm{\mu} =  \begin{bmatrix} \mu_{1} \\[0.3em] \mu_{2} \end{bmatrix}, 
	\quad \mathbf{B} = 
	\begin{bmatrix}
		\mathbf{b}_{1,1}^\intercal  & \mathbf{b}_{1,2}^\intercal  \\[0.3em]
		\mathbf{b}_{2,1}^\intercal  & \mathbf{b}_{2,2}^\intercal
	\end{bmatrix}_{2 \times (p_1 + p_2)} \; \text{and}
	\quad
	\mathbf{X}_t = \begin{bmatrix} X_{t,1} \\[0.3em] X_{t,2} \end{bmatrix}  \in \mathbb{R}^{p_1 + p_2}.
\end{equation*}
Indeed, the bivariate CARMA(p,q)-Hawkes process reads
\begin{equation}\label{eq:IntensityPCHD}
\bm{\lambda}_t = \bm{\mu} + \mathbf{B} \mathbf{X}_t 
\end{equation}
and 
\begin{equation*}
\mathrm{d} \mathbf{X}_t  = \bar{\mathbf{A}} \mathbf{X}_t \mathrm{d}t + \bar{\mathbf{e}}\mathrm{d}\mathbf{N}_t, \; \text{with} \; X_0 = 0,
\end{equation*}
where
\begin{equation*}\label{ematr}
\bar{\mathbf{A}} = 
 \begin{bmatrix}
  \mathbf{A}_{1}  & 0  \\[0.3em]
               0  & \mathbf{A}_{2}
 \end{bmatrix}
\quad 
\text{and} \ \ 
\bar{\mathbf{e}} = 
 \begin{bmatrix}
  \mathbf{e}_{1}  & 0  \\[0.3em]
               0  & \mathbf{e}_{2}
 \end{bmatrix}_{(p_1 + p_2) \times 2}.
\end{equation*}
We remark that the state process of $\bm{\lambda}_t$ in \eqref{eq:IntensityPCHD} is still a linear stochastic differential equation where the driving noise is a bivariate counting process and has an analytical solution.



\noindent The likelihood of the bivariate CARMA(p,q)-Hawkes process, where $\theta = \left(\bm{\mu}, \bar{\mathbf{A}}, \mathbf{B} \right)$ contains the set of parameters to estimate, reads
\begin{equation}\label{eq: lf1}
\mathcal{L}\left(\theta\right)  = \displaystyle \sum_{i = 1}^{2} \left[ - \int_{0}^{T} \lambda_{t, i} \mathrm{d}t + \int_{0}^{T} \ln \left(\lambda_{t, i} \right) \mathrm{d}N_{t, i} \right].
\end{equation}
Let us denote $\mathbbm{1} =  \left[1 \hspace{0.4em} 1 \right]^{\top}$ and $\ln \left(\bm{\lambda}_t \right)  =  \left[\ln \left(\lambda_{t, 1} \right) \hspace{0.4em} \ln \left(\lambda_{t, 2} \right) \right]^{\top}$, then \eqref{eq: lf1} can be rewritten as 
\begin{equation}\label{eq: lf2}
\mathcal{L}\left(\theta\right) =  - \int_{0}^{T} \mathbbm{1}^\intercal \bm{\lambda}_{t} \mathrm{d}t + \int_{0}^{T} \ln \left(\bm{\lambda}_{t} \right)^\intercal \mathrm{d}\mathbf{N}_{t}.
\end{equation}
Substituting \eqref{eq:IntensityPCHD} in \eqref{eq: lf2} and making some calculations, we finally get
\begin{equation}
	\begin{split}
\mathcal{L} \left(\theta\right) & =  - \int_{0}^{T} \mathbbm{1}^\intercal \left[\bm{\mu} + \mathbf{B} \mathbf{X}_t \right] \mathrm{d}t + \int_{0}^{T} \ln \left(\bm{\mu} + \mathbf{B} \mathbf{X}_t \right)^\intercal \mathrm{d}\mathbf{N}_{t}\\
&= - \mathbbm{1}^\intercal \bm{\mu} \left(T - 0 \right) - \mathbbm{1}^\intercal \mathbf{B} \int_{0}^{T} \mathbf{X}_t \mathrm{d}t + \displaystyle \sum_{T_k \leq T}   \ln \left(\bm{\mu} + \mathbf{B} \mathbf{X}_{T_k} \right)^\intercal \Delta \mathbf{N}_{T_k}  \label{likelihood}
	\end{split}
\end{equation}
where $\Delta\mathbf{N}_{T_k}: = \mathbf{N}_{T_n} - \mathbf{N}_{T_k-}$ and $\mathbf{N}_{T_k-}$ is the left limit of the bivariate process. Note that $\Delta\mathbf{N}_{T_k} =  \left[1 \hspace{0.4em} 0 \right]^{\top}$ (or equivalently $\left[0 \hspace{0.4em} 1 \right]^{\top}$) 
where $T_k$ is the bivariate jump time process; specifically, $T_k =  \left[T_{k, 1} \hspace{0.4em} T_{k, 2} \right]^{\top}$.

\noindent Recalling  
that  $\bar{\mathbf{e}}$ is a $(p_1+p_2)\times 2$ matrix 
with zero elements, except for $\mathbf{e}_{p_1, 1}=1$ and $\mathbf{e}_{(p_1+p_2), 2}=1$, we consider the following summation defined iteratively as
\begin{equation*}
\begin{split}
s(1) &=  \bar{\mathbf{e}}\Delta \mathbf{N}_{T_1}  \\
s(n) &=  \bar{\mathbf{e}}\Delta \mathbf{N}_{T_n} + e^{\bar{\mathbf{A}} \left(T_n - T_{n-1} \right)}s(n-1).
\end{split}
\end{equation*}
Assuming that $T = T_n$ (i.e., we have $n$ jumps in $[0, T_n]$),  
the likelihood function of the bivariate CARMA(p,q)-Hawkes finally becomes
\begin{equation}
\mathcal{L}\left(\theta\right) =  - \mathbbm{1}^\intercal \bm{\mu}T_n + \mathbbm{1}^\intercal \mathbf{B}\bar{\mathbf{A}}^{-1} \bar{\mathbf{e}} \mathbf{N}_{T_n} - \mathbbm{1}^\intercal \mathbf{B}\bar{\mathbf{A}}^{-1}  s(n)  + \displaystyle \sum_{i = 1}^{n}   \ln \left(\bm{\mu} + \mathbf{B} \mathbf{X}_{T_i} \right)^\intercal \Delta \mathbf{N}_{T_i}.
\label{likefin}
\end{equation}
See \ref{appendice} for the derivation if \eqref{likefin} . 

\section{Description of the estimation procedure}\label{estimation}
The estimation exercise presented in the paper is based on an iterative sequential procedure which considers three different frameworks: the bivariate CARMA(p,q)-Hawkes  (bCH) framework applied directly to the series of upward and downward price movements, the univariate CARMA(p,q)-Hawkes combined with the Lee and Mykland test for jump detection (hereafter uCHLM framework), and the bivariate CARMA(p,q)-Hawkes with Lee and Mykland test (bCHLM framework). We first discuss the test of Lee and Mykland (hereafter LM test) proposed in \cite{lee2008jumps} and then dive into the details of the considered frameworks. 

\noindent Let $P(t_i)$ be the asset price at time $t_i$ for $i=1,\ldots,n$. In the LM test, the quantity $\mathcal{M}(i)$, is defined as
\begin{equation}\label{lmtest}
	\mathcal{M}(i) := \frac{\log \frac{P(t_i)}{P(t_{i-1})}}{\hat{\sigma}(t_i)},
\end{equation}
with $\hat{\sigma}^2(t_i)$ given by
\begin{equation*}
	\hat{\sigma}^2(t_i) := \frac{1}{K-2}\sum_{j=i-K+2}^{i-1}|\log \frac{P(t_j)}{P(t_{j-1})}||\log \frac{P(t_{j-1})}{P(t_{j-2})}|,
\end{equation*} 
where $K$ is the window size. Note that \eqref{lmtest} tests at time $t_i$ if a jump happened in the time interval from $t_{i-1}$ to $t_i$. Here we state the following asymptotic result based on \cite[Lemma 1]{lee2008jumps}, assuming zero drift for the dynamics of $P_t$ in but in the \textsl{highfrequency} \texttt{R} package described in \cite{boudt2022} the LM test is implemented also in the case with possible non zero drift term. 

\noindent Let be $K = O_p(\Delta t^{\bar{\alpha}})$ where $-1 < \bar{\alpha} < -0.5$ and assume zero drift for the dynamics of $P_t$, then $\text{as} \quad \Delta t \rightarrow 0$
\begin{equation}
	\frac{\mathcal{M}(i)-C_n}{S_n}\rightarrow \eta 
	\label{testspec}
\end{equation} 
where $\eta$ has a cumulative distribution function $P(\eta\leq x)=\exp(-e^{-x})$. The quantities $C_n$ and $S_n$ are respectively 
\begin{equation*} 
	C_n=\frac{(2\log n)^{1/2}}{c}-\frac{\log \pi +\log (\log n)}{2c(2\log n)^{1/2}} \ \ \text{and} \ \ S_n=\frac{1}{c(2 \log n)^{1/2}},
\end{equation*}
where $n$ denotes the number of observations and $c\approx 0.7979$. In this paper we use the LM test for detecting the jumps within an intraday data set, but other tests can be considered for the detection of jumps. For instance, a list of non-parametric tests for detecting jumps presented in \cite{dumitru2012identifying} can be used in order to extend the proposed analyses.

\subsection{Frameworks under investigation}\label{modelsjd}



\noindent Here we present the frameworks under investigation and we recall that we look for the autoregressive and moving average orders of CARMA(p,q)-Hawkes models (univariate or bivariate). The main difference of the frameworks under investigation refers to the data we give as input to the model to be fitted, i.e. the definition of jumps. The three frameworks are sequential since if one fails to be accepted (see Section~\ref{estimation_procedure} for the type of test), then the algorithm moves on to the next one. 
\begin{enumerate}
	\item \textbf{The bCH framework} In this framework any price movement is used to define $N^{+}_T$ and $N^{-}_T$ that refer respectively to the sequence of positive and negative price movements (i.e.,  $\mathbf{N}_t=\left[N^{+}_{T} \hspace{0.4em} N^{-}_{T}\right]^{\top}$ is a bivariate point process where the intensities are defined in \eqref{eq:IntensityPCHD}). 
	\item \textbf{The uCHLM framework}. 
	Here, we consider the univariate CARMA(p,q)-Hawkes model fitted to the univariate vector $N_T$ that contains jumps which in this context refer to the abnormal price movements. Jumps are identified by means of the LM test. Four different levels of confidence intervals for the LM test are considered, namely \newline $\alpha_{LM}\in \{95\%, 97.5\%, 99\%, 99.5\%\}$.
	
	\item \textbf{The bCHLM framework}. Similar to the bCH framework, but, in this case, we consider the vector of jumps detected through the LM test by distinguishing between positive and negative jumps. In practice, we repeat the whole procedure applied to the newly defined vector $\mathbf{N}_t=\left[N^{+}_{T} \hspace{0.4em} N^{-}_{T}\right]^{\top}$ for $\alpha_{LM}\in \{95\%, 97.5\%, 99\%, 99.5\%\}$.
\end{enumerate}

\subsection{Routine of the estimation procedure}\label{estimation_procedure}
\begin{figure}[!htbp]
\centering
\begin{tikzpicture}[scale=0.99] 
	\path \etape{1}{Data set};
	\path (p1.south)+(0, -1.4) \etape{2}{Framework model: \\ the routine starts from the simplest one};

	\path (p1.south)+(-3,-3.7) \etape{3}{\textbf{Candidate model}: \\ Choosing $\textbf{p}$ and \textbf{q}};

	\path (p3.south)+(0,-1.2) \etape{4}{Model estimation};
	\path (p1.south)+(3,-3.7) \etape{5}{\textbf{Alternative model}: \\ Choosing $\textbf{p}$ and \textbf{q}};
	\path (p5.south)+(0,-1.2) \etape{6}{Model estimation};
	
	\path (p4.south)+(0,-1.4) \etape{7}{Optimal parameters for candidate model};
	\path (p6.south)+(0,-1.4) \etape{8}{Optimal parameters for alternative model};

    \path (p2.south)+(-2.5,-7.8) \etape{10}{LR test for nested models};
    \path (p2.south)+(2.5,-7.8) \etape{11}{AIC selection criterion for non-nested models};
    
	\draw[arrow] (p1) -- (p2);
	\draw[arrow] (p3) -- (p4);
	\draw[arrow] (p5) -- (p6);

	\draw [arrow] (p6.south) --  node[above] {}(p8.north) ;
	\draw [arrow] (p4.south) --  node[above] {}(p7.north) ;

	\begin{scope}[on background layer]
		\node (byellow) [draw, thick, draw=yellow!50!black, fill=yellow!20, rounded corners, fit=(p3) (p4) (p5) (p6), inner xsep=5pt, inner ysep=6pt,  label={[node font=\normalsize\bfseries,rotate=0,anchor=south, xshift=33mm, yshift=-1.5mm]north: Estimation phase}] {};

		\node (bred) [draw, thick, red!50!black, fill=red!75!black!25, rounded corners, fit=(p10) (p11), inner xsep=5pt, inner ysep=6pt,  label={[node font=\normalsize\bfseries,rotate=0,anchor=south, xshift=30mm, yshift=-0.5mm]: Model selection}] {};
	\end{scope}

   \path (bred.south)+(-105pt,-62pt)  \decision{14}{Candidate model selected and there is at least an alternative model nearby};
   \path (bred.south)+(-75pt,-113.5pt)  \decision{15}{Candidate model selected and there is no alternative model};
   \path (bred.south)+(75pt,-113.5pt)  \decision{16}{Alternative model selected and there are no models nearby};
   \path (bred.south)+(105pt,-62pt)  \decision{17}{Alternative model selected and there is at least a model nearby};

\begin{scope}[on background layer]

		\node (bgreen) [draw, thick, green!50!black, fill=green!75!black!25, rounded corners, fit=(p14) (p15) (p16) (p17), inner xsep=5pt, inner ysep=6pt,  label={[node font=\normalsize\bfseries,rotate=0,anchor=south, xshift=44mm, yshift=-0.5mm]: Scenarios}] {}
%

;
\end{scope}

   \path (p2.south)+(0.0,-14.9) \etape{18}{Residual analysis performed by KS test};
   
   \path (p18.south)+(0.0,-1.6) \etape{19}{Best selected model};

	\draw [arrow] (p7.east) -|  node[above] {}([xshift=-1.5mm, yshift=0mm]bred.north) ;
	\draw [arrow] (p8.west) -|  node[above] {}([xshift=1.5mm, yshift=0mm]bred.north) ;

	\draw [arrow] (bred.south) --  node[above] {}(p14.north) ;
	\draw [arrow] (bred.south) --  node[above] {}(p17.north) ;
	
	\draw [arrow] (bred.south) --  node[above] {}([xshift=15.5mm, yshift=0mm]p15.north) ;
	\draw [arrow] (bred.south) --  node[above] {}([xshift=-15.5mm, yshift=0mm]p16.north) ;
	
	\draw [arrow] (p15) --  node[above] {}(p18.north) ;
	\draw [arrow] (p16) --  node[above] {}(p18.north) ;
	
	\draw [arrow] (p18) --  node[text width= 6.5cm,  xshift=35mm] {Not rejecting $H_0$}(p19.north) ;
	
	\draw[arrow] ([xshift=-17.9mm]p14.north) --  node [text width=7cm,above, rotate=90,  xshift=2.5mm] {Candidate mod. remains candidate mod.} ([xshift=-2.4mm]byellow.west) -- ([xshift=0mm, yshift=0mm]byellow.west);
	
	\draw[arrow] ([xshift=17.9mm]p17.north) --  node [text width=7.5cm,above, rotate=270,  xshift=-0.5mm] {Alternative mod. becomes candidate mod.} ([xshift=2.4mm]byellow.east) -- ([xshift=0mm, yshift=0mm]byellow.east);
	
	\draw[arrow] (p2) -- (byellow);
	
	\draw[arrow]  (p18.west) node[text width= 6.5cm, below,  xshift=0mm] {Rejecting $H_0$} -- ([xshift=-45mm]p18.west) node [text width=6.5cm,above, rotate=90,  xshift=123mm ] {Moving to the next framework model} |- (p2.west);
\end{tikzpicture}
\caption{Description for the analysis of the microstructure for a given time series of bond prices. The three alternative frameworks are 1) The bivariate CARMA(p,q)-Hawkes applied to jumps identified at any price movement 2) The Univariate CARMA(p,q)-Hawkes model applied to all the jumps identified through the LM test and 3) The bivariate CARMA(p,q)-Hawkes applied to positive and negative jumps identified through the LM test. \label{Procfig}}
\end{figure}
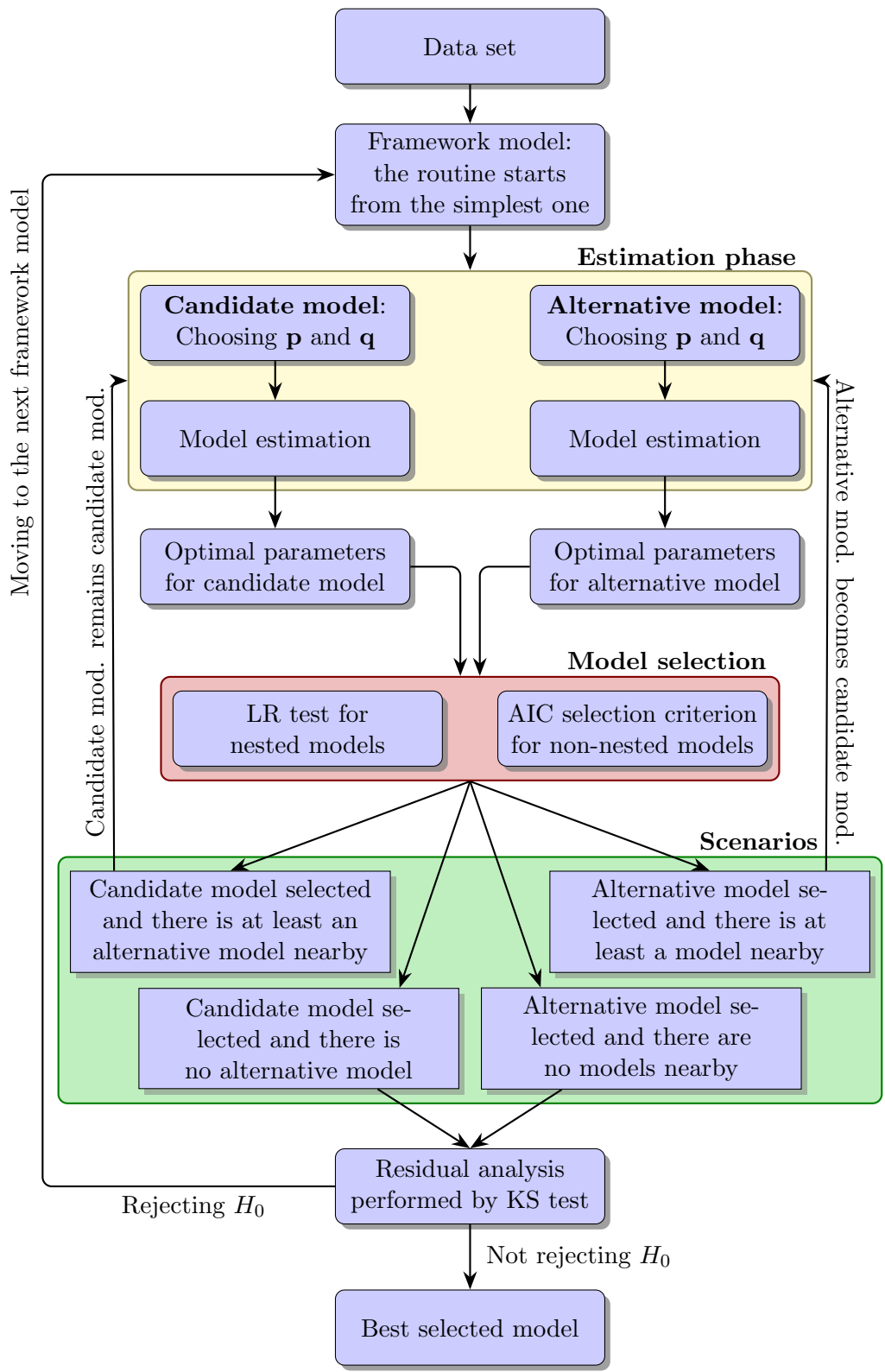

\noindent The estimation procedure for the analysis starts from the simplest model (i.e., the bCH framework); see the diagram in Figure \ref{Procfig}. Then it proceeds with the estimation phase (yellow box) in which two models are fitted: the candidate and the alternative models. Such estimation is based on the maximization of the likelihood function. Once obtained the optimal parameters, a model selection between the candidate and alternative models is performed (red box). 

In case of non-nested models the model selection is exploits  the Akaike information criterion (AIC) while in case of nested models, model selection is  based on the likelihood-ratio (LR) test. In traditional statistical theory, such a test is widely accepted for testing the goodness of fit between nested models.
The statistic of the likelihood ratio test, i.e., $LR$, can be computed as the difference of two loglikelihoods in nested models. Specifically,
\begin{equation}
	LR = 2 \left(\log \mathcal{L}^{\text{unrest.}}\left(\theta\right) -   \log \mathcal{L}^{\text{rest.}}\left(\theta\right)\right).
\end{equation}
 is asymptotically distributed as a $\chi^2$ with degrees of freedom equal to the number of additional parameters in the unrestricted model (see \cite{kendall1977} for further details). For sake of clarity, we provide an example. Suppose that we choose for the candidate model (bCH framework), which is the restricted one, the following orders $\mathbf{p}  = \left[1,1 \right]$ and $\mathbf{q}  = \left[0, 0, 0, 0 \right]$. For the alternative model, which is the unrestricted one, a natural choice is to consider the model with $\mathbf{p} = \left[2, 1\right]$ and $\mathbf{q} = \left[1,0,1,0\right]$. In case of rejection of the null hypothesis, we consider a new unrestricted model\footnote{At this stage the alternative model becomes the candidate model.} (i.e., a new alternative model); for instance, a larger combination of orders such  as $\mathbf{p} = \left[2, 2\right]$ and $\mathbf{q} = \left[1,1,1,1\right]$. Thus the idea, in case of rejection of the null hypothesis, is keep testing nearby models until the null hypothesis fails to be rejected. Note that, considering nested and non-nested model, the routine generates four possible scenarios (see green box of Figure \ref{Procfig}). Two scenarios refer to the possibility that there is at least an alternative or nearby model to verify while the other two concern the case in which there are no alternative or nearby models. If that is the situation, then the optimal orders of that framework are identified.  
After the identification of the best fitting model, the algorithm moves on to the next step which is the \textsl{residual analysis}. Such an analysis is used to establish whether the data are generated by the selected model and works as follows.
Let us suppose that the observed time arrivals $\left\{t_i\right\}_{i=1,\ldots,n}$ are generated by the conditional intensity $\lambda_t$. Then we construct a new set of time arrivals $\left\{\tau_i\right\}_{i=1,\ldots,n}$, as follows:
\begin{equation}
	\tau_i:=\int_{0}^{t_i}\lambda_t \mbox{d}t.
\end{equation}
The transformed data $\tau_i$ are expected to follow a stationary Poisson process with unitary intensity (see for example \cite{brown2002time} for details) and can be used to test the form of the conditional intensity. In our routine we implement the Kolmogorov-Smirnov (KS) test, which tests the residual differences between the empirical and theoretical cumulative distribution functions of the inter-event (or inter-arrival) times. Specifically,  inter-event times are denoted as
\begin{equation*}
	u_i := \tau_i-\tau_{i-1},
\end{equation*} 
with $\tau_0=0$. If $\lambda_t$ is correctly specified (i.e., $\lambda_t$ is positive over $[0, T]$ and $\int_{0}^{T}\lambda_t \mbox{d}t<\infty$ a.s.), then $\left\{u_i\right\}_{i=1,\ldots,n}$ are i.i.d. exponentially distributed with unit rate.
If the identified model  does not pass the residual analysis, then the algorithm moves on to the next framework. If the opposite happens, then the procedure identifies the best model and stops. Figure \ref{Procfig} displays an overview of the routine.

%
%
%
%

\section{Empirical Analysis}
\label{empirical}
\subsection{Descriptive analysis}
In this section we discuss the results of the estimation and the selection procedure previously described conducted on different bonds issued by Cr\'edit Agricole SA (financial institution issuing green bonds since 2013) and Engie SA (corporate company operating in the energy market, the largest non sovereign green bond issuer in France) observed in two different time periods.  The main assumption is that in summer and in autumn the news in the energy market are quite relevant for highlighting differences in green and brown bonds. Moreover, we deal with two time frameworks where there is a difference based on whether the ECB made interest rate announcements or not. The data refer to tick data (ask and bid prices) downloaded from Bloomberg where we consider the Green label indicator. The first period ranges from August 19 to September 19, 2022 while the second period ranges from November 14 to December, 2022.  For Cr\'edit Agricole we consider two types of bonds (bullet and callable) while for Engie only bullet bonds. We select pairs of bonds (green and brown) with similar maturity, coupon rate, currency and rating.\newline
\noindent The main features of these instruments are reported in Table \ref{BondsCA} while in Tables \ref{mainstat} and \ref{mainstat2} we present the main statistics of the bid-ask spread for the three groups (green and brown bonds) for two high frequency data with time window length equal to one month.  It results that the observed bid-ask spread of green bonds on average is larger for green bonds issued by Cr\'edit Agricole SA , especially in the second time period. The opposite happens for the bonds issued by Engie SA. For the callable bonds issued by Cr\'edit Agricole SA  the results are mixed. Indeed, the currency of these bond is the Singapore Dollar (SGD) and there are quotations only for the half trading day. It clearly emerges that some intraday news are absorbed with a different time lag with respect to the series of bonds quoted in the European market. We also highlight the fact that bid-ask spreads for the bullet bonds (green and brown) do not display significant differences in the two non-overlapping time intervals while for the case of the callable bonds we observe on average larger values and standard deviation in the case of the green instrument. 
     
\begin{table}[!htbp]
\centering
\begin{small}
\begin{tabular}{llllllll}
\hline\hline
\multicolumn{8}{c}{\textbf{Bullet}}\\
\textbf{Coupon} & \textbf{Maturity}   & \textbf{Series} & \textbf{Rating}  & \textbf{Announce}   & \textbf{Currency} & \textbf{Ask Price} & \textbf{Type}\\
\hline
0.375  & 21/10/2025 & EMTN   & A-   & 14/10/2019  & EUR & 92.611   & Green    \\
0.5    & 24/06/2024 & EMTN   & A-   & 17/06/2019  & EUR & 96.344   & Brown    \\
\hline\hline
\multicolumn{8}{c}{\textbf{Callable}}\\
\textbf{Coupon} & \textbf{Maturity}   & \textbf{Series} & \textbf{Rating}  & \textbf{Announce}   & \textbf{Currency} & \textbf{Ask Price} & \textbf{Type}\\
\hline
3.95 &	22/07/2032	& EMTN	& BBB+	&	13/04/2022	& SGD &	97.697	&	Green \\
3.8	 & 30/04/2031	  & EMTN	& BBB+	&	23/04/2019	& SGD	& 98.364	&	Brown \\
\hline\hline
\multicolumn{8}{c}{\textbf{Bullet Corporate}}\\
\textbf{Coupon} & \textbf{Maturity}   & \textbf{Series} & \textbf{Rating}  & \textbf{Announce}   & \textbf{Currency} & \textbf{Ask Price} & \textbf{Type}\\
\hline
0.875 &	27/03/2024	& EMTN	& BBB+	&	15/03/2017	& EUR &	97.481	&	Green \\
0.375	 & 11/06/2027	  & EMTN	& BBB+	&	04/06/2020	& EUR	& 87.163	&	Brown \\
\hline\hline
\end{tabular}
\caption{Features of bonds included in the dataset. The issuers are Cr\'edit Agricole SA  (financial) and Engie SA (corporate). Prices refer to September 19, 2022.\label{BondsCA}}
\end{small}
\end{table}

\begin{table}[!htbp]
\begin{small}
\begin{tabular}{lccccccccc}
\hline \hline
\textbf{Aug-Sep}     & \textbf{Mean}   & \textbf{Med.} & \textbf{Mode}   & \textbf{Std.} & \textbf{Ex. Kur.} & \textbf{Skew.} & \textbf{IQR}  & \textbf{Min} & \textbf{Max} \\
\hline
\textbf{Bullet  Fin(G)}    & 0.2650 & 0.2640 & 0.2560 & 0.0113             & 0.1904   & 0.1945   & 0.1080 & 0.2230  & 0.3310 \\
\textbf{Bullet Fin (B)}    & 0.1987 & 0.1970 & 0.1910 & 0.0111             & 0.2999   & 0.5853   & 0.1070 & 0.1630  & 0.2700  \\
\textbf{Callable Fin (G)}  & 0.8193 & 0.8180 & 0.7650 & 0.2350             & -0.7506  & 0.3930   & 0.2300 & 0.7210  & 0.9510  \\
\textbf{Callable Fin (B)}  & 0.8869 & 0.8230 & 0.7880 & 0.2342             & 12.6908  & 3.6924   & 1.7220 & 0.7470  & 2.4690 \\
\textbf{Bullet Corp (G)}    & 0.2374 & 0.2390 & 0.2440 & 0.0302             & -0.8252   & -0.0418   & 0.0491 & 0.1710  & 0.350 \\
\textbf{Bullet Corp (B)}    & 0.3677 & 0.3660 & 0.3360 & 0.0279             & -0.3826   & 0.6439   & 0.0430 & 0.2990  & 0.4460  \\
\hline \hline
\end{tabular}
\end{small}
\caption{Main statistics of the bid-ask spread for the bonds included in the dataset. The label in brackets refers to the bond's type \textsl{G} for green and \textsl{B} for brown. The data refer to the period August 19 - September 19, 2022.\label{mainstat}}
\end{table}

\begin{table}[!htbp]
\begin{small}
\begin{tabular}{lccccccccc}
\hline \hline
\textbf{Nov-Dec}     & \textbf{Mean}   & \textbf{Med.} & \textbf{Mode}   & \textbf{Std.} & \textbf{Ex. Kur.} & \textbf{Skew.} & \textbf{IQR}  & \textbf{Min} & \textbf{Max} \\
\hline
\textbf{Bullet Fin (G)}    & 0.2869 & 0.2860 & 0.2830 & 0.0094 & 0.6992   & 0.4862   & 0.0770 & 0.2580  & 0.3350  \\
 \textbf{Bullet Fin (B)} & 0.1854 & 0.1830 & 0.1790 & 0.0102 & -0.2050  & 0.5818   & 0.0660 & 0.1560  & 0.2220  \\
\textbf{Callable Fin (G)} & 2.9652 & 3.0610 & 3.4090 & 0.3837 & 0.2112   & -0.8960  & 1.4180 & 2.0890  & 3.5070  \\
\textbf{Callable Fin (B)} & 1.5423 & 1.5000 & 1.4990 & 0.1841 & -1.0681  & -0.0598  & 0.6380 & 1.2010  & 1.8390 \\
\textbf{Bullet Corp (G)}    & 0.2709 & 0.2680 & 0.2620 & 0.0191             & 0.2792   & 0.461   & 0.0260 & 0.2130  & 0.3441 \\
\textbf{Bullet Corp (B)}    & 0.3463 & 0.3610 & 0.3690 & 0.0212             & -0.8068   & -0.1887   & 0.0360 & 0.3090  & 0.4270  \\
\hline \hline
\end{tabular}
\end{small}
\caption{Main statistics of the bid-ask spread for the bonds included in the dataset. The label in brackets refers to the bond's type \textsl{G} for green and \textsl{B} for brown. The data refer to the period November 14 - December 13, 2022.\label{mainstat2}}
\end{table}
\clearpage

\subsection{Analysis of the microstructure based on jumps}
\begin{figure}[!htbp]
	\begin{subfigure}[b]{0.5\textwidth}
	\centering
		\includegraphics[width=\textwidth]{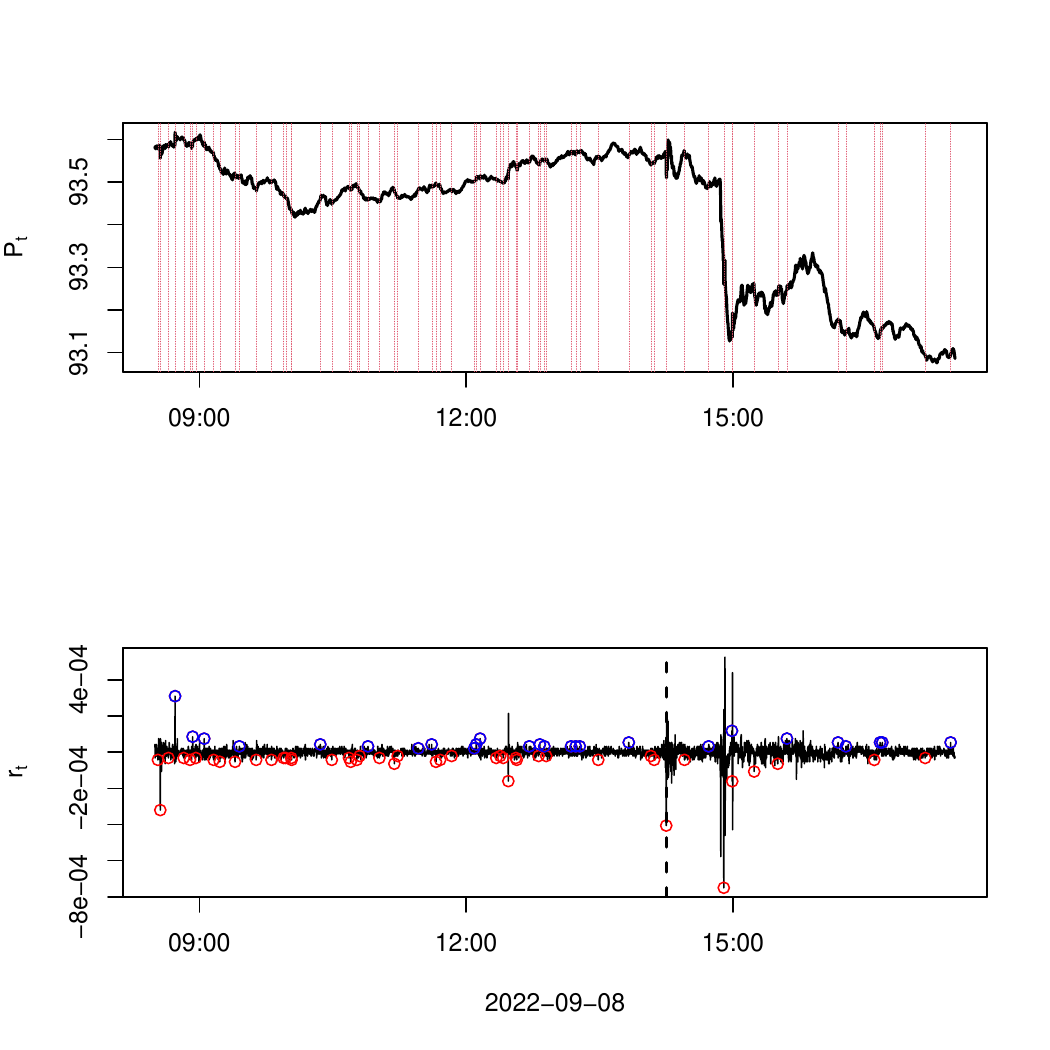}
	\end{subfigure}
     \hfill
     \begin{subfigure}[b]{0.5\textwidth}
		\includegraphics[width=\textwidth]{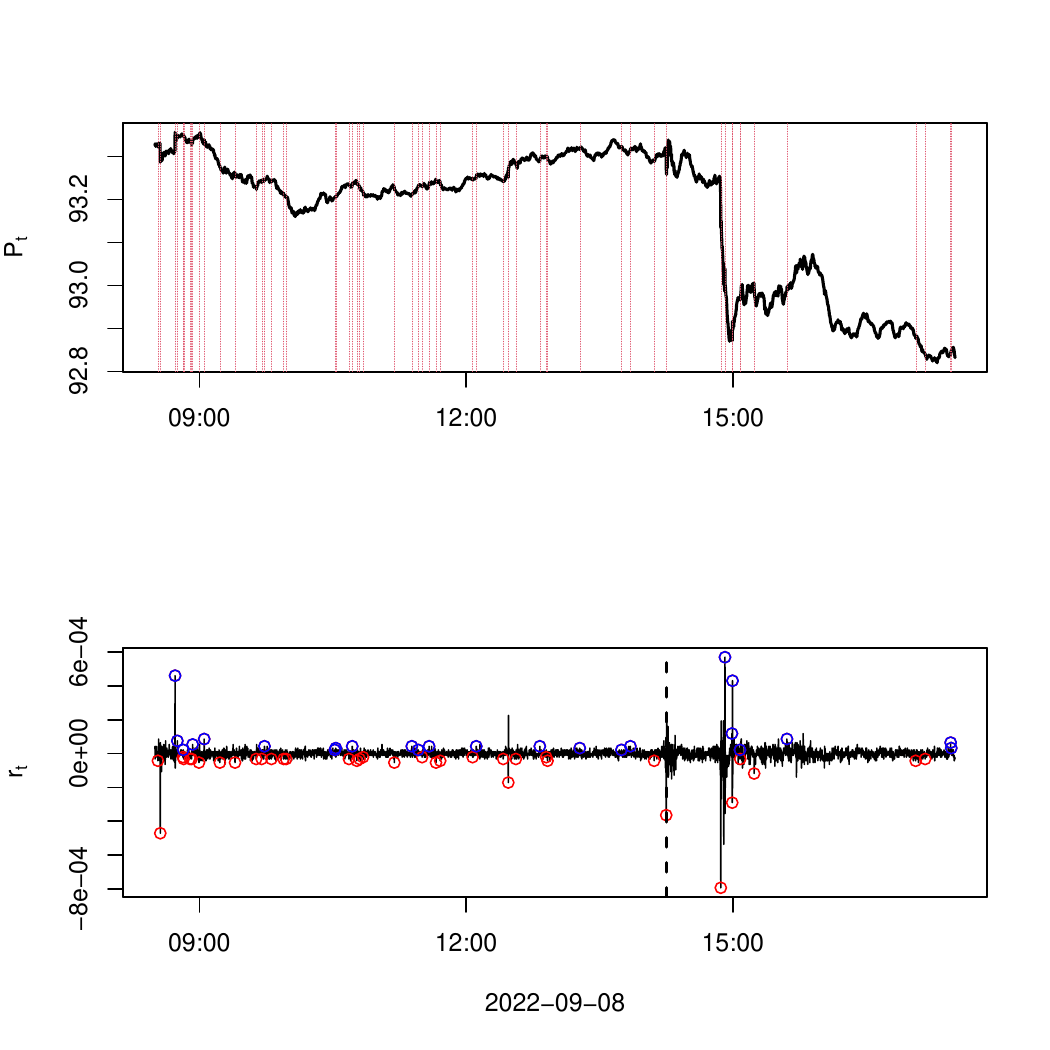}
	\end{subfigure}
	\caption{Prices and returns of the green bullet bond issued by Cr\'edit Agricole SA  observed at September 8, 2022 (plots on the left refer to ask prices/returns while plots on the right to the bid counterparts). Red and blue circles refer respectively to negative and positive jumps in the series of returns identified through the LM test for $\alpha_{LM}=0.95$. Vertical red lines in the upper plots refer to time instants where (positive or negative) jumps occur. \label{fig:ds_old_ACAFP_0_375_10_21_2025_GB}}
\end{figure}

\begin{figure}[!htbp]
	\begin{subfigure}[b]{0.5\textwidth}
	\centering
		\includegraphics[width=\textwidth]{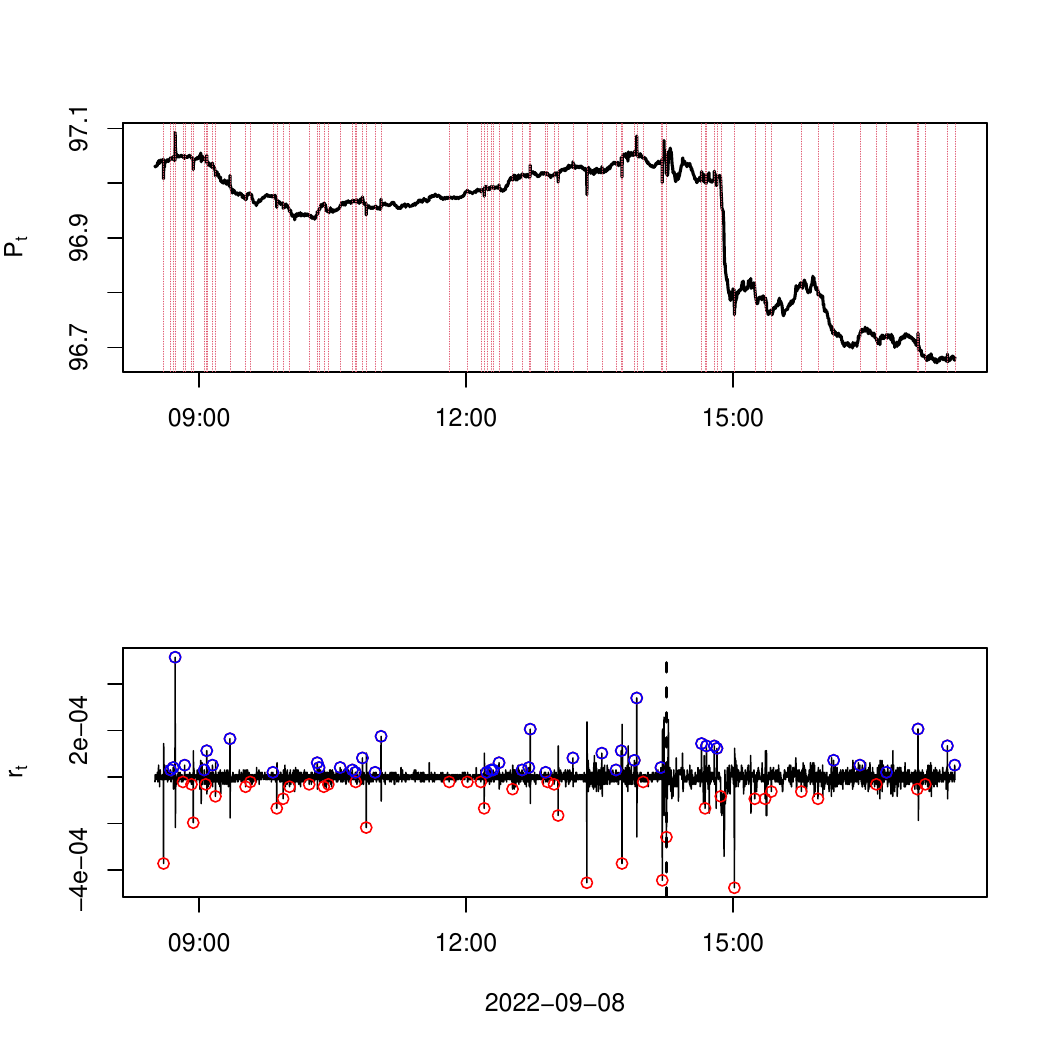}
	\end{subfigure}
     \hfill
     \begin{subfigure}[b]{0.5\textwidth}
		\includegraphics[width=\textwidth]{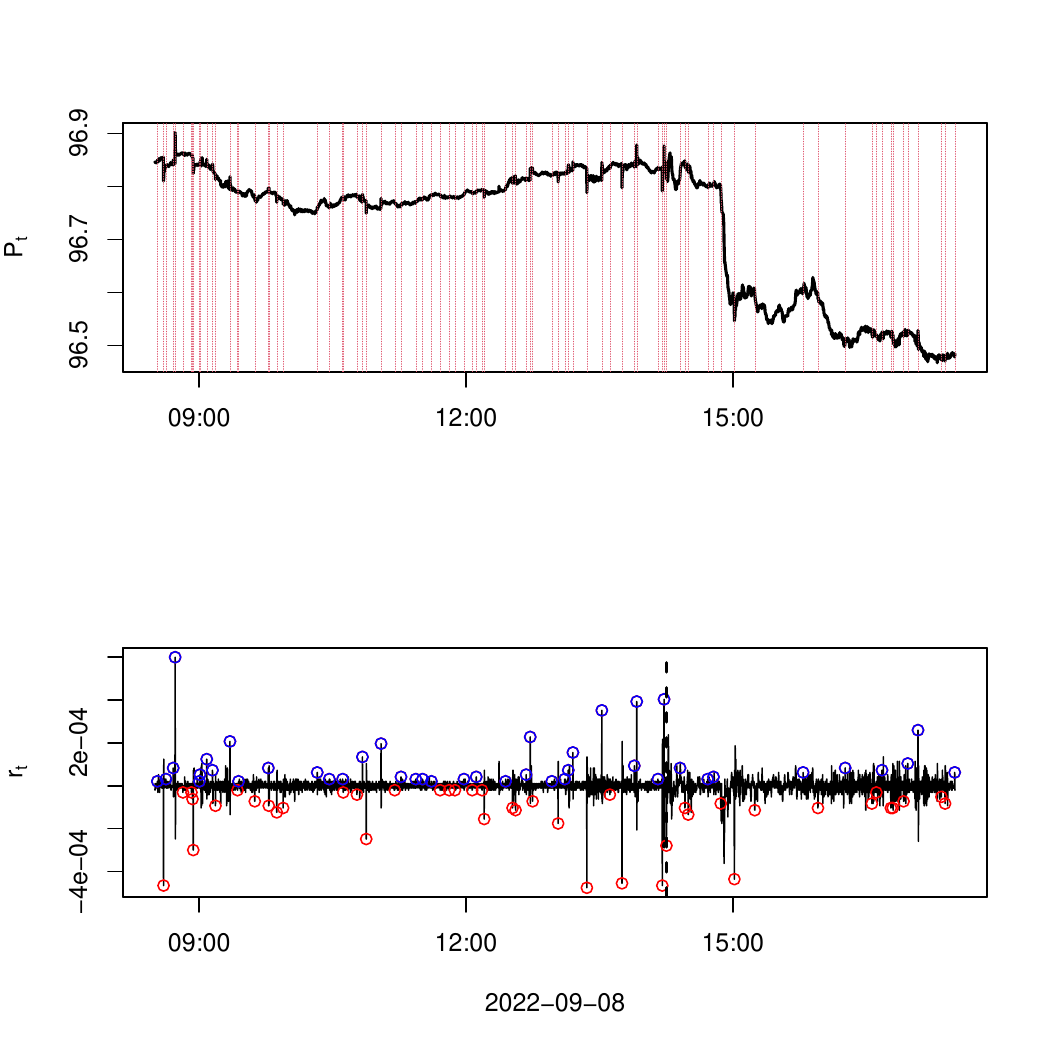}
	\end{subfigure}
	\caption{Prices and returns of the brown bullet bond issued by Cr\'edit Agricole SA  observed at September 8, 2022 (plots on the left refer to ask prices/returns while plots on the right to the bid counterparts). Red and blue circles refer respectively to negative and positive jumps in the series of returns identified through the LM test for $\alpha_{LM}=0.95$. Vertical red lines in the upper plots refer to time instants where (positive or negative) jumps occur.  \label{fig:ds_old_ACAFP_0_5_2024_BB}}
\end{figure}

\begin{figure}[!htbp]
	\begin{subfigure}[b]{0.5\textwidth}
	\centering
		\includegraphics[width=\textwidth]{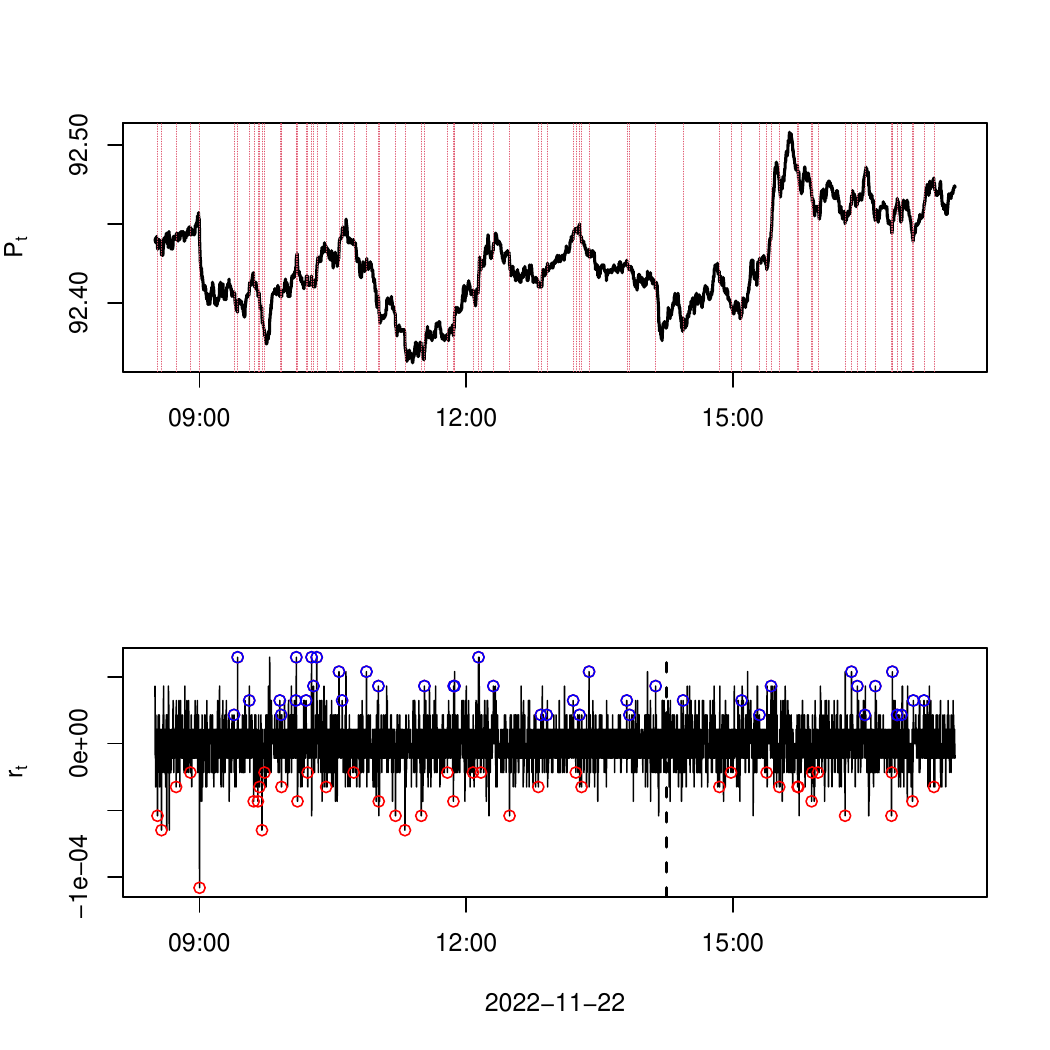}
	\end{subfigure}
     \hfill
     \begin{subfigure}[b]{0.5\textwidth}
		\includegraphics[width=\textwidth]{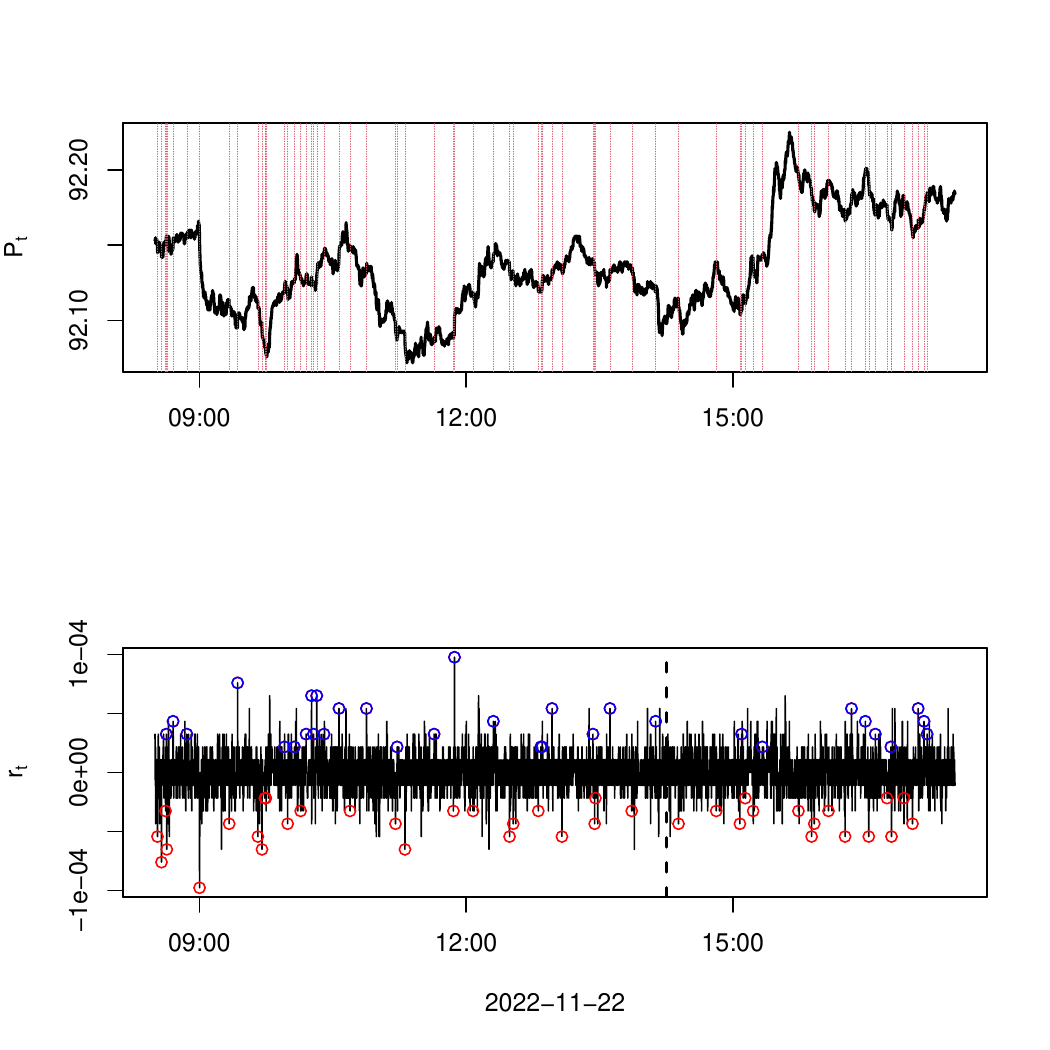}
	\end{subfigure}
	\caption{Prices and returns of the green bullet bond issued by Cr\'edit Agricole SA  observed at November 22, 2022 (plots on the left refer to ask prices/returns while plots on the right to the bid counterparts). Red and blue circles refer respectively to negative and positive jumps in the series of returns identified through the LM test for $\alpha_{LM}=0.95$. Vertical red lines in the upper plots refer to time instants where (positive or negative) jumps occur. \label{fig:ds_rec_ACAFP_0_375_10_21_2025_GB}}
\end{figure}

\begin{figure}[!htbp]
	\begin{subfigure}[b]{0.5\textwidth}
	\centering
		\includegraphics[width=\textwidth]{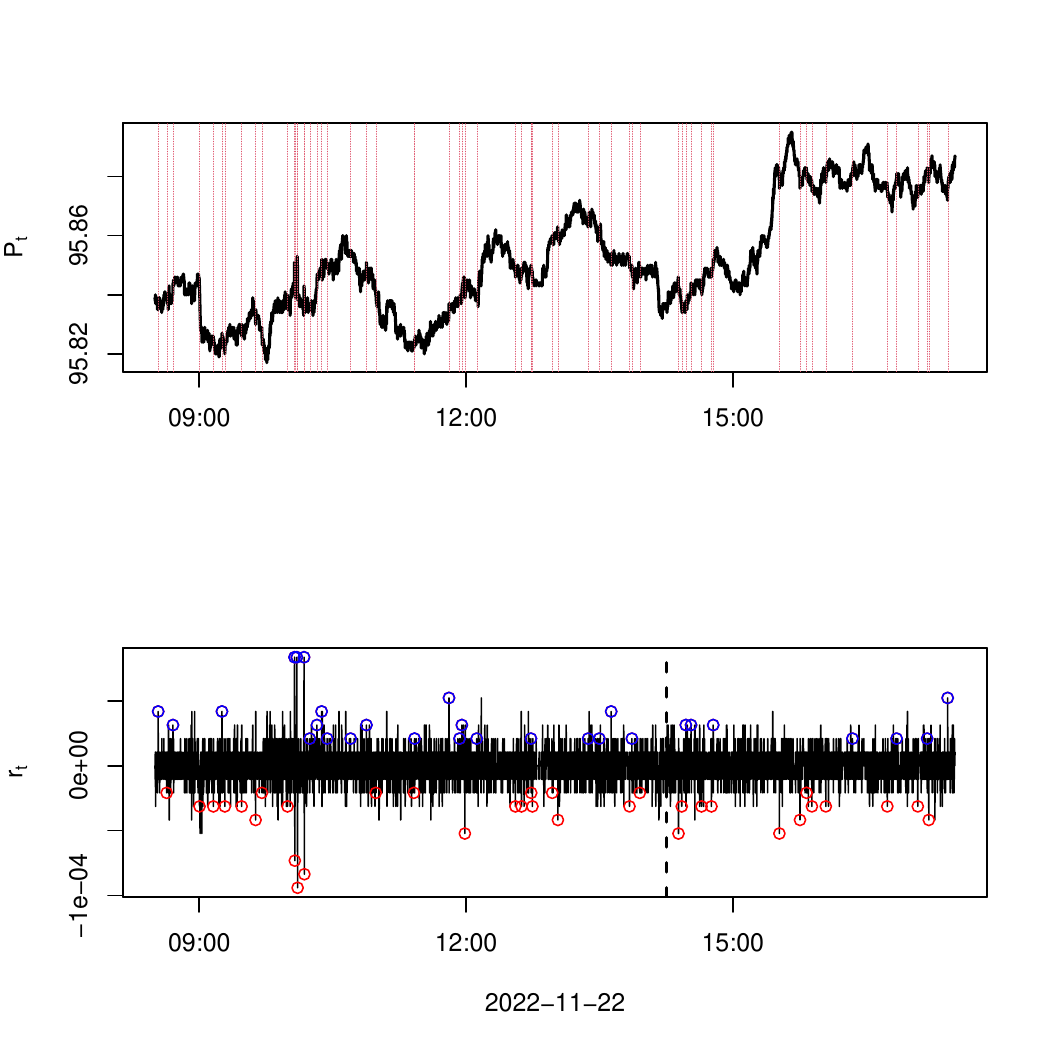}
	\end{subfigure}
     \hfill
     \begin{subfigure}[b]{0.5\textwidth}
		\includegraphics[width=\textwidth]{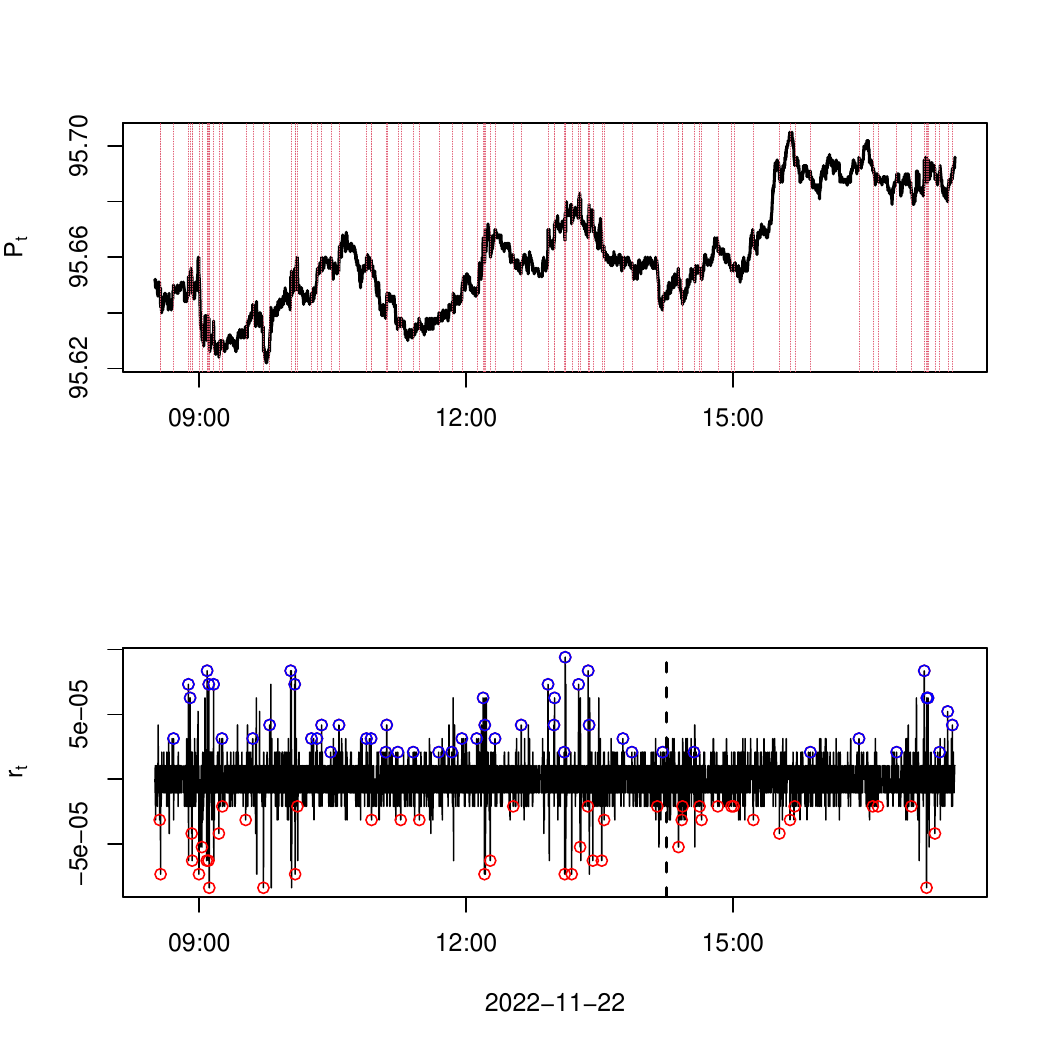}
	\end{subfigure}
	\caption{Prices and returns of the brown bullet bond issued by Cr\'edit Agricole SA  observed at November 22, 2022 (left ask; right bid). Red and blue circles refer respectively to negative and positive jumps in the series of returns identified through the LM test for $\alpha_{LM}=0.95$. Vertical red lines in the upper plots refer to time instants where (positive or negative) jumps occur. \label{fig:ds_rec_ACAFP_0_5_2024_BB}}
\end{figure}

\begin{figure}[htbp]
	\begin{subfigure}[b]{0.5\textwidth}
	\centering
		\includegraphics[width=\textwidth]{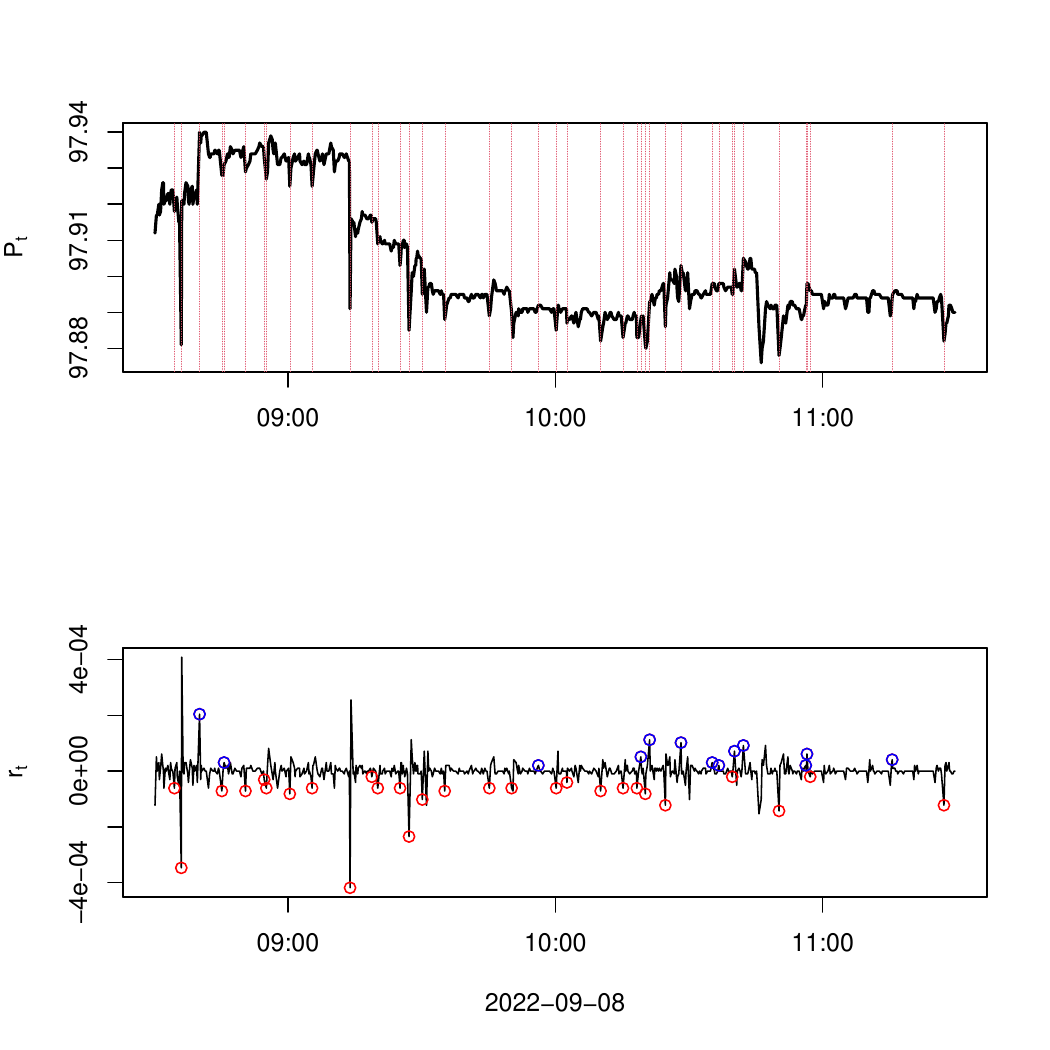}
	\end{subfigure}
     \hfill
     \begin{subfigure}[b]{0.5\textwidth}
		\includegraphics[width=\textwidth]{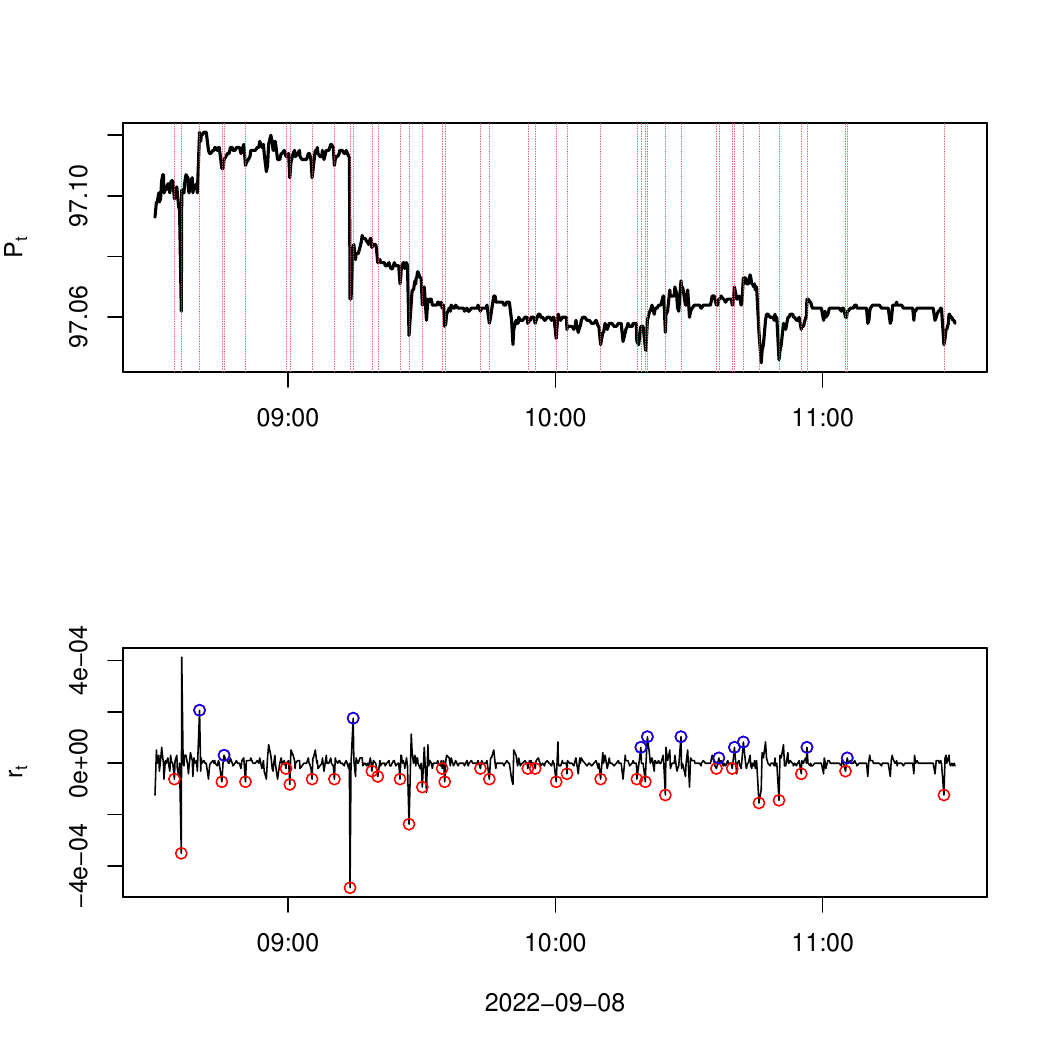}
	\end{subfigure}
	\caption{Prices and returns of the green callable bond issued by Cr\'edit Agricole SA  observed at September 8, 2022 (left ask; right bid). Red and blue circles refer respectively to negative and positive jumps in the series of returns identified through the LM test for $\alpha_{LM}=0.95$. Vertical red lines in the upper plots refer to time instants where (positive or negative) jumps occur. \label{fig:ds_old_ACAFP_3_95_2032_GB_CA_todo_M}}
\end{figure}

\begin{figure}[htbp]
	\begin{subfigure}[b]{0.5\textwidth}
	\centering
		\includegraphics[width=\textwidth]{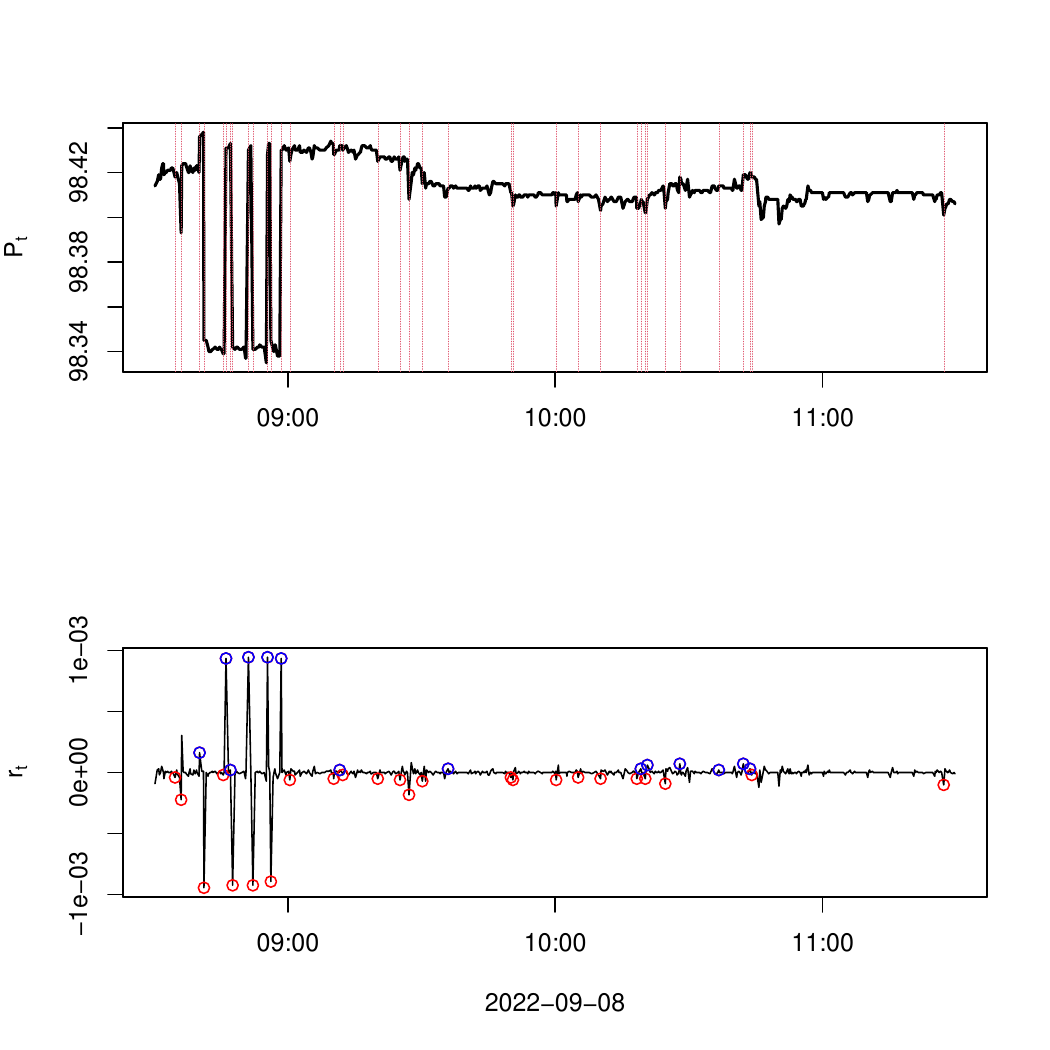}
	\end{subfigure}
     \hfill
     \begin{subfigure}[b]{0.5\textwidth}
		\includegraphics[width=\textwidth]{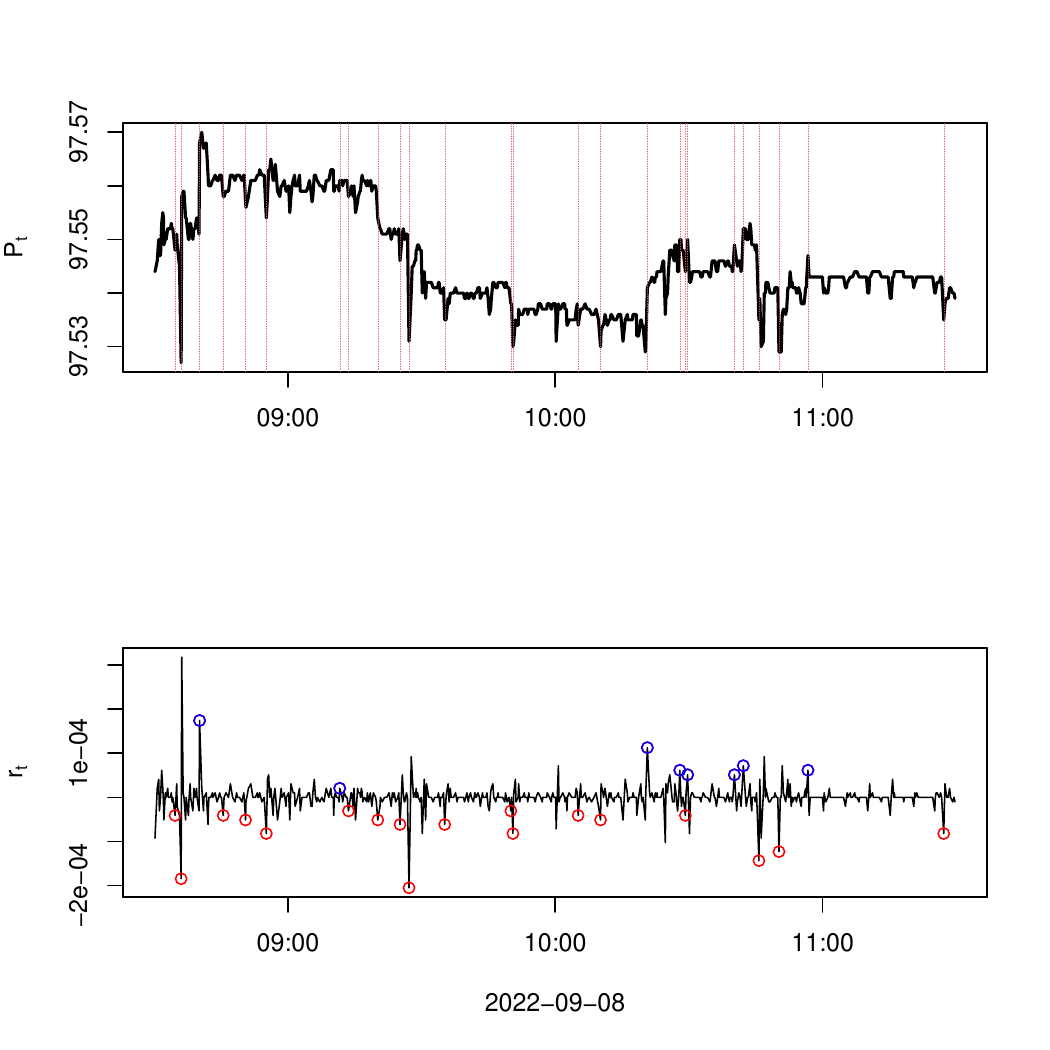}
	\end{subfigure}
	\caption{Prices and returns of the brown callable bond issued by Cr\'edit Agricole SA  observed at September 8, 2022 (left ask; right bid). Red and blue circles refer respectively to negative and positive jumps in the series of returns identified through the LM test for $\alpha_{LM}=0.95$. Vertical red lines in the upper plots refer to time instants where (positive or negative) jumps occur. \label{fig:ds_old_ACAFP_3_8_2031_BB_CA}}
\end{figure}

\begin{figure}[htbp]
	\begin{subfigure}[b]{0.5\textwidth}
	\centering
		\includegraphics[width=\textwidth]{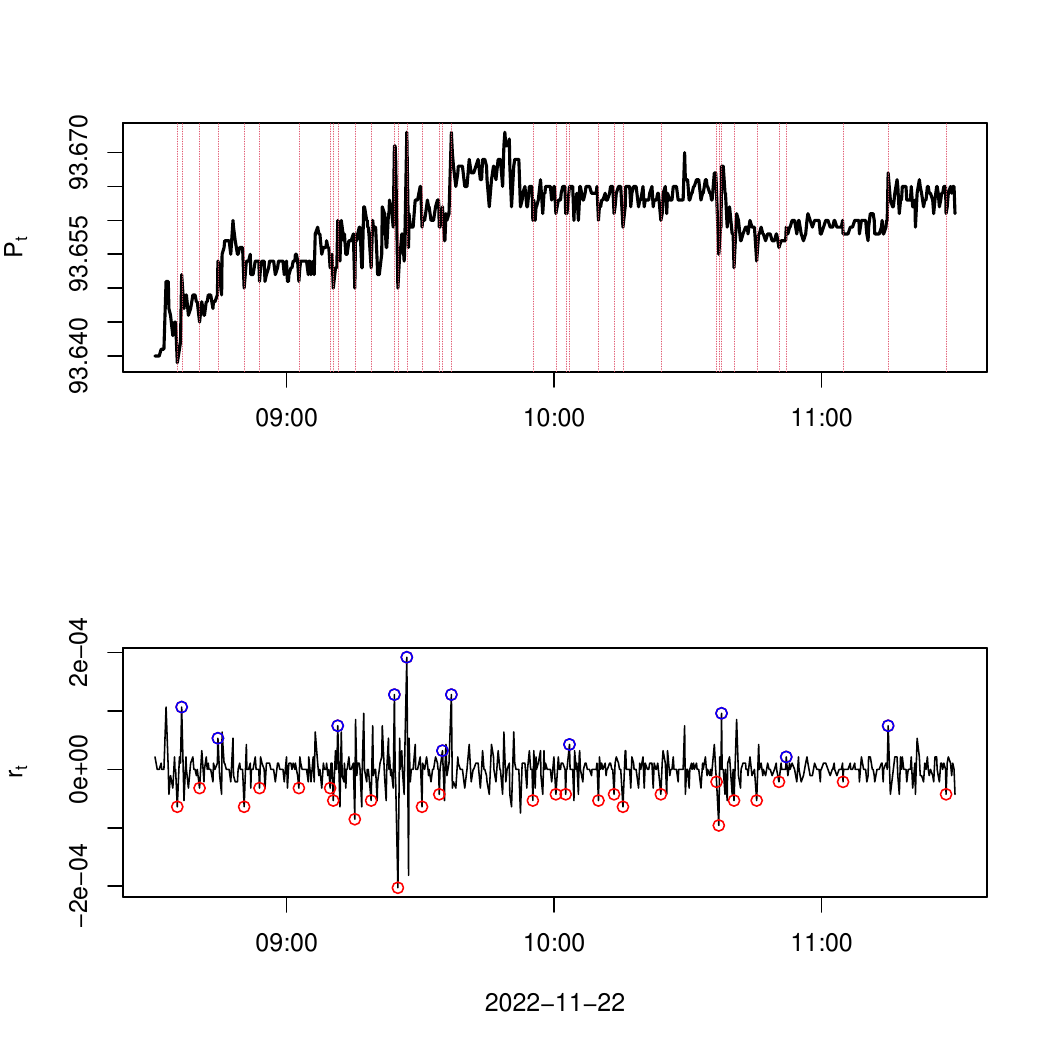}
	\end{subfigure}
     \hfill
     \begin{subfigure}[b]{0.5\textwidth}
		\includegraphics[width=\textwidth]{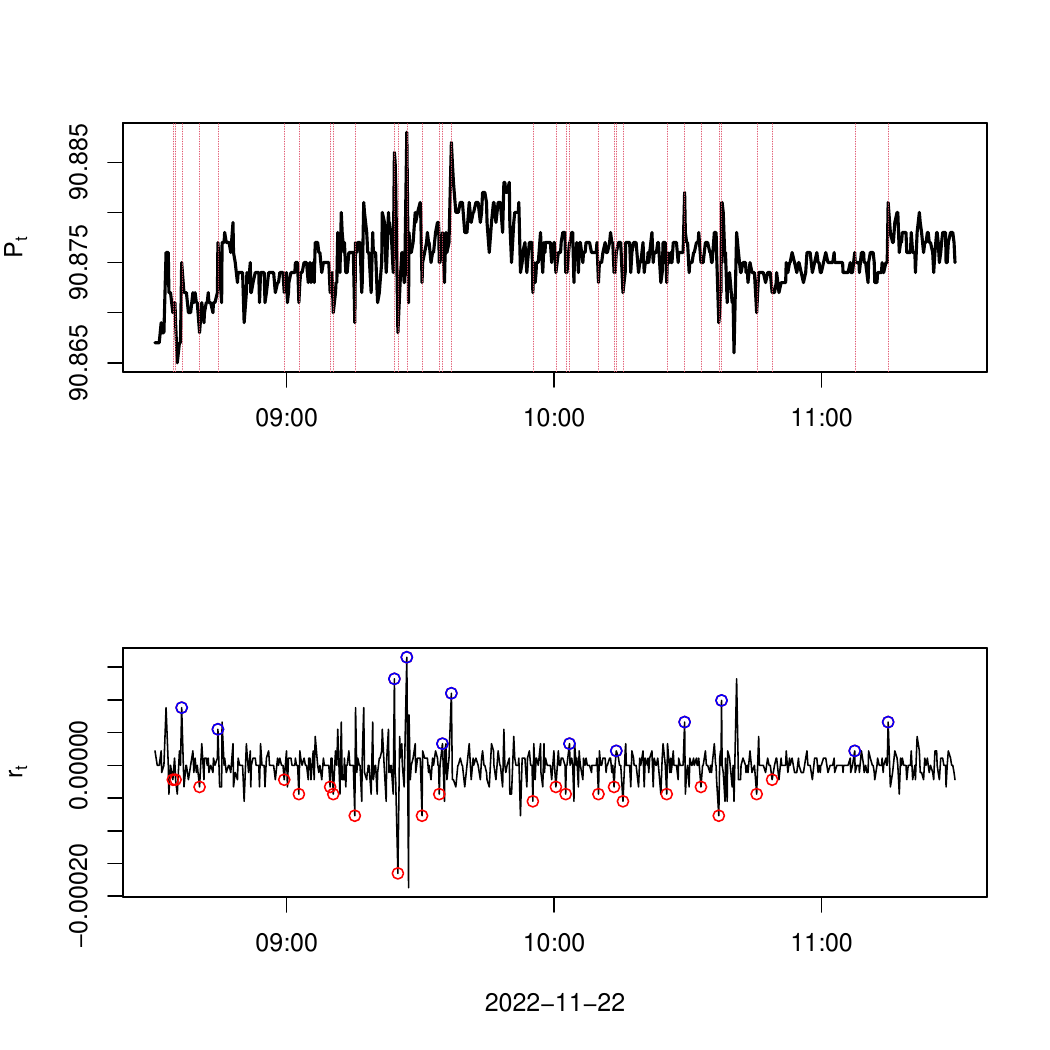}
	\end{subfigure}
	\caption{Prices and returns of the green callable bond issued by Cr\'edit Agricole SA  observed at November 22, 2022 (left ask; right bid). Red and blue circles refer respectively to negative and positive jumps in the series of returns identified through the LM test for $\alpha_{LM}=0.95$. Vertical red lines in the upper plots refer to time instants where (positive or negative) jumps occur. \label{fig:ds_rec_ACAFP_3_95_2032_GB_CA_todo_M}}
\end{figure}

\begin{figure}[htbp]
	\begin{subfigure}[b]{0.5\textwidth}
	\centering
		\includegraphics[width=\textwidth]{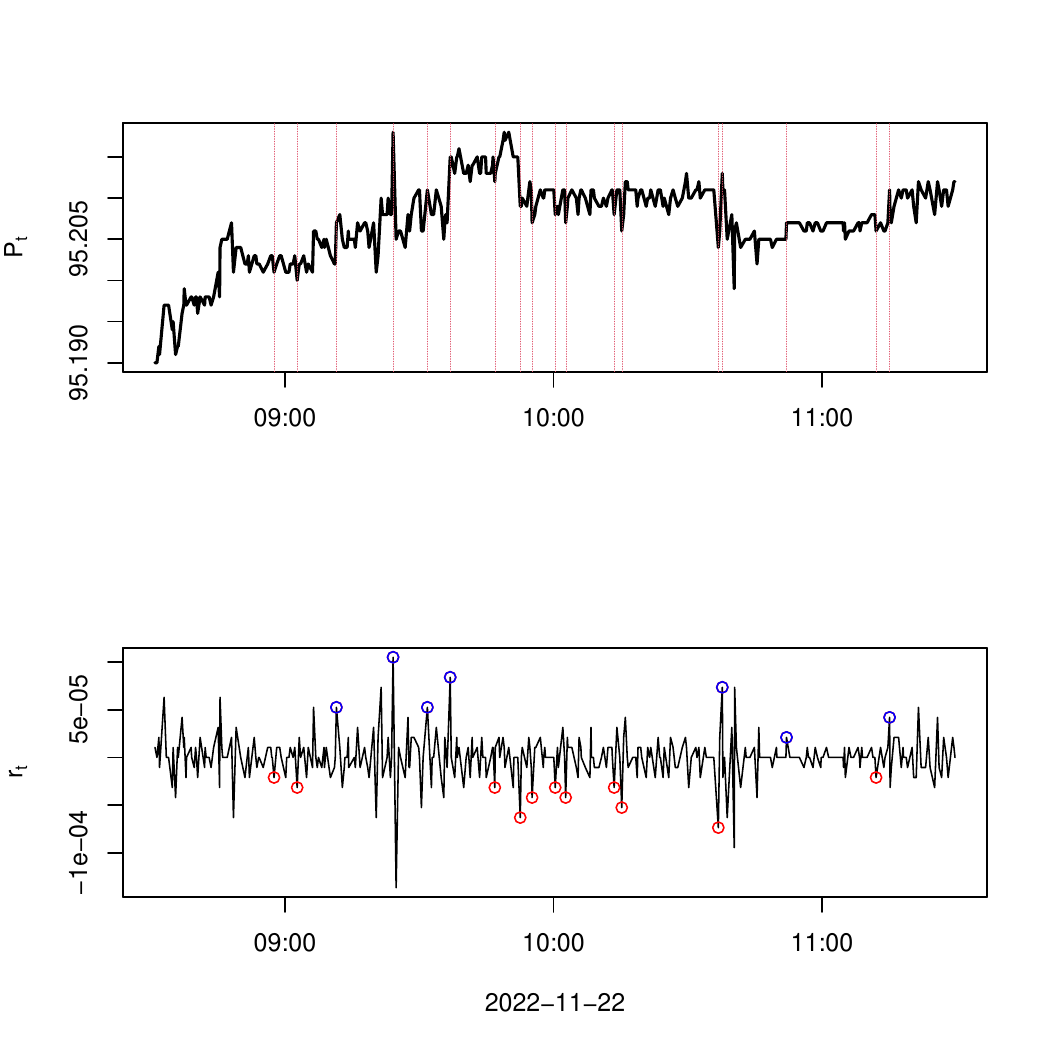}
	\end{subfigure}
     \hfill
     \begin{subfigure}[b]{0.5\textwidth}
		\includegraphics[width=\textwidth]{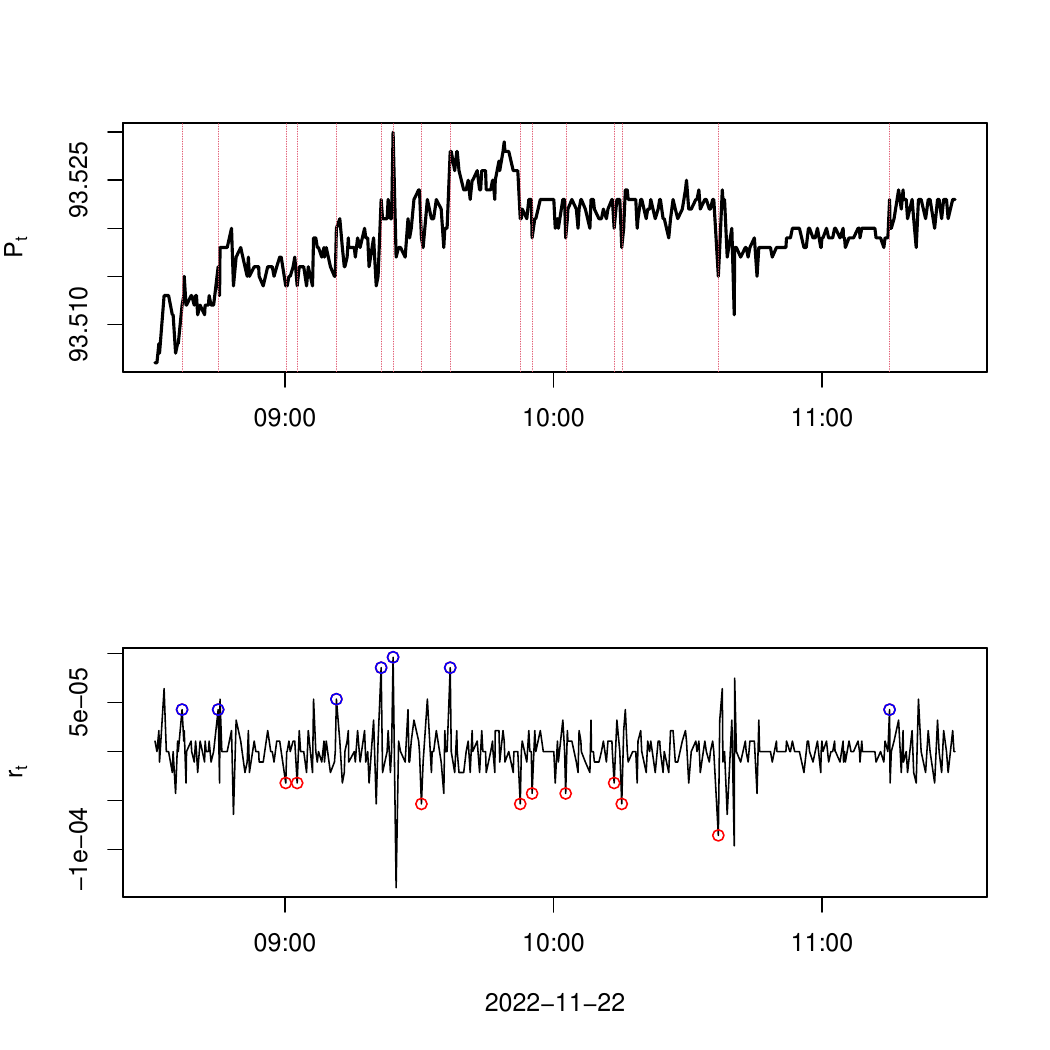}
	\end{subfigure}
	\caption{Prices and returns of the brown callable bond issued by Cr\'edit Agricole SA  observed at November 22, 2022 (plots on the left refer to ask prices/returns while plots on the right to the bid counterparts). Red and blue circles refer respectively to negative and positive jumps in the series of returns identified through the LM test for $\alpha_{LM}=0.95$. Vertical red lines in the upper plots refer to time instants where (positive or negative) jumps occur. \label{fig:ds_rec_ACAFP_3_8_2031_BB_CA}}
\end{figure}

\begin{figure}[htbp]
	\begin{subfigure}[b]{0.5\textwidth}
		\centering
		\includegraphics[width=\textwidth]{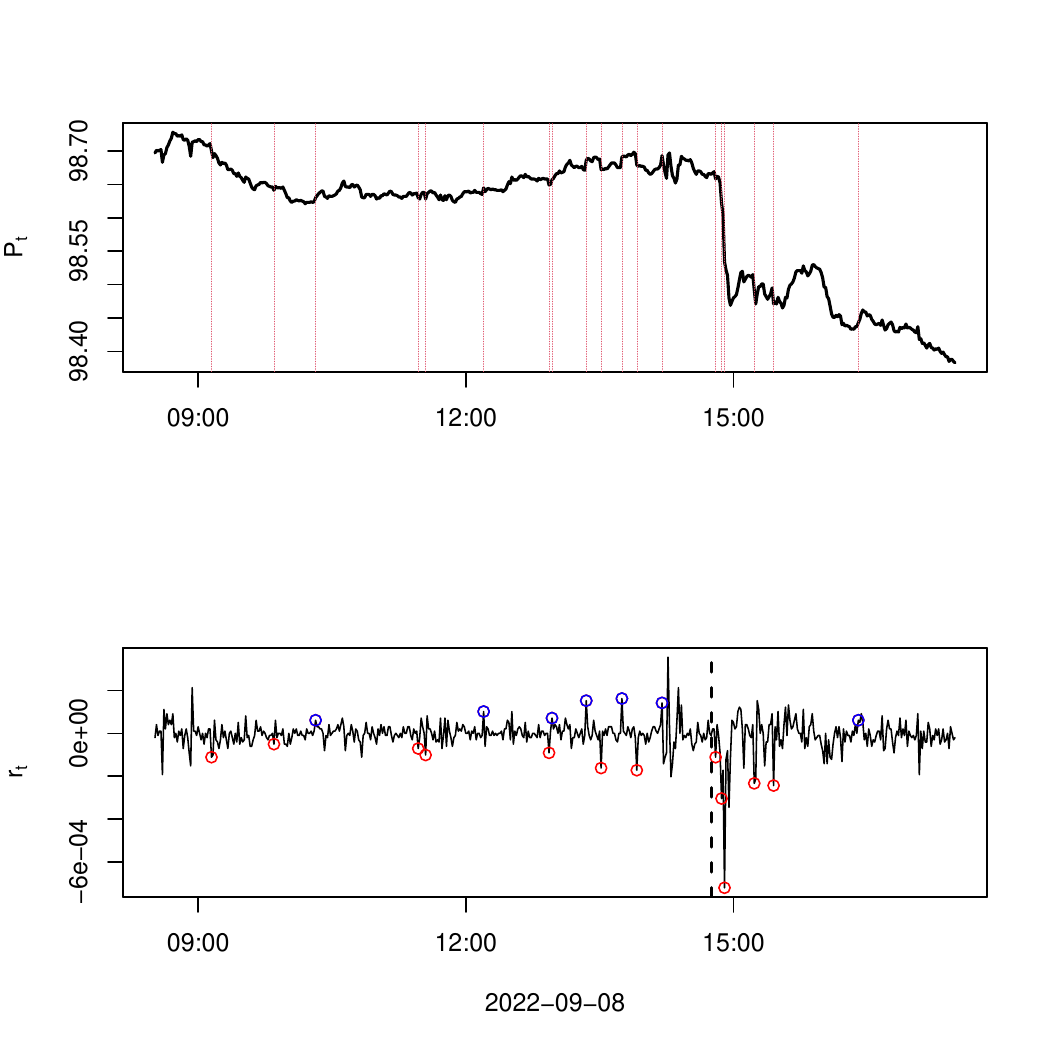}
	\end{subfigure}
	\hfill
	\begin{subfigure}[b]{0.5\textwidth}
		\includegraphics[width=\textwidth]{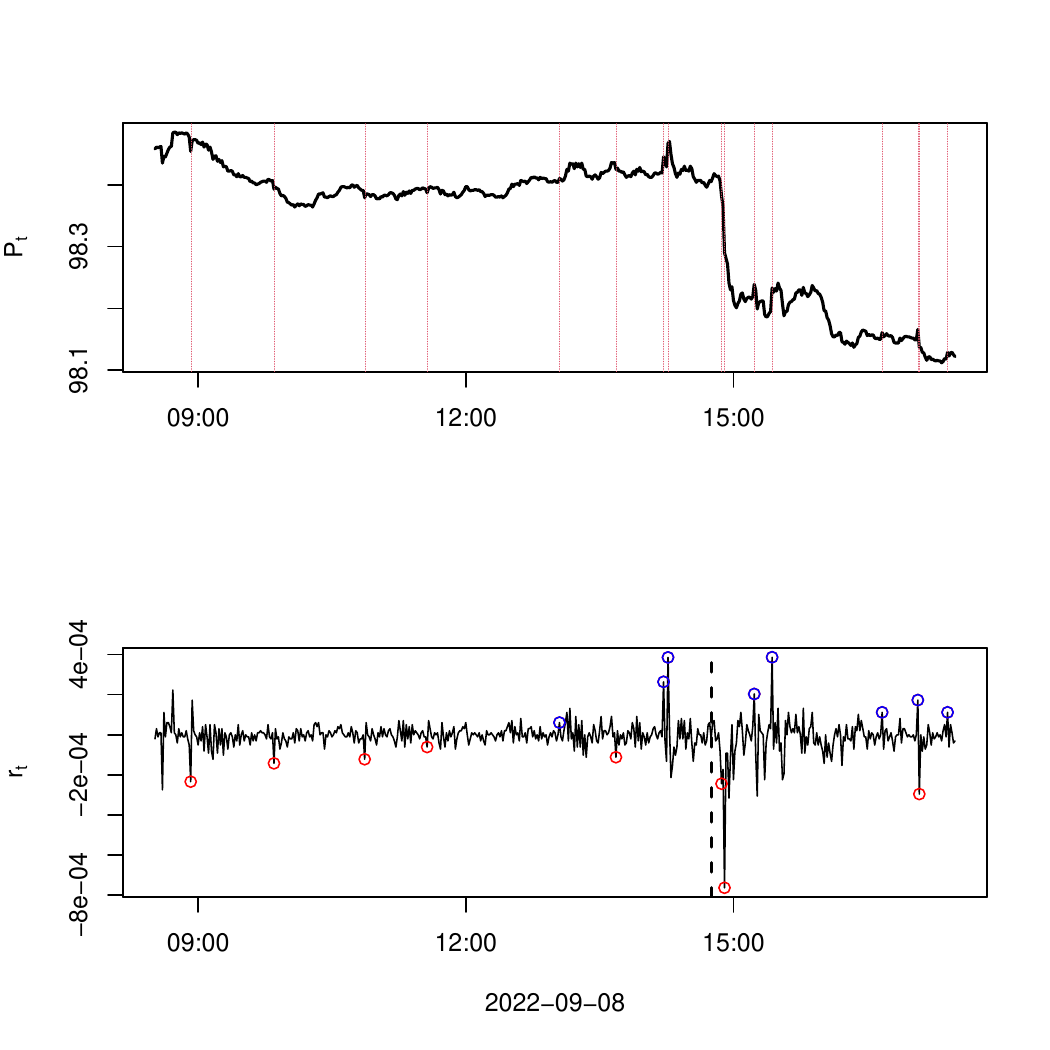}
	\end{subfigure}
	\caption{Prices and returns of the green bullet bond issued by Engie SA observed observed at September 8, 2022 (plots on the left refer to ask prices/returns while plots on the right to the bid counterparts). Red and blue circles refer respectively to negative and positive jumps in the series of returns identified through the LM test for $\alpha_{LM}=0.95$. Vertical red lines in the upper plots refer to time instants where (positive or negative) jumps occur. \label{fig:ds_ENGIFP_0_875_2024_corp_green_OLD_RIGHT}}
\end{figure}

\begin{figure}[htbp]
	\begin{subfigure}[b]{0.5\textwidth}
		\centering
		\includegraphics[width=\textwidth]{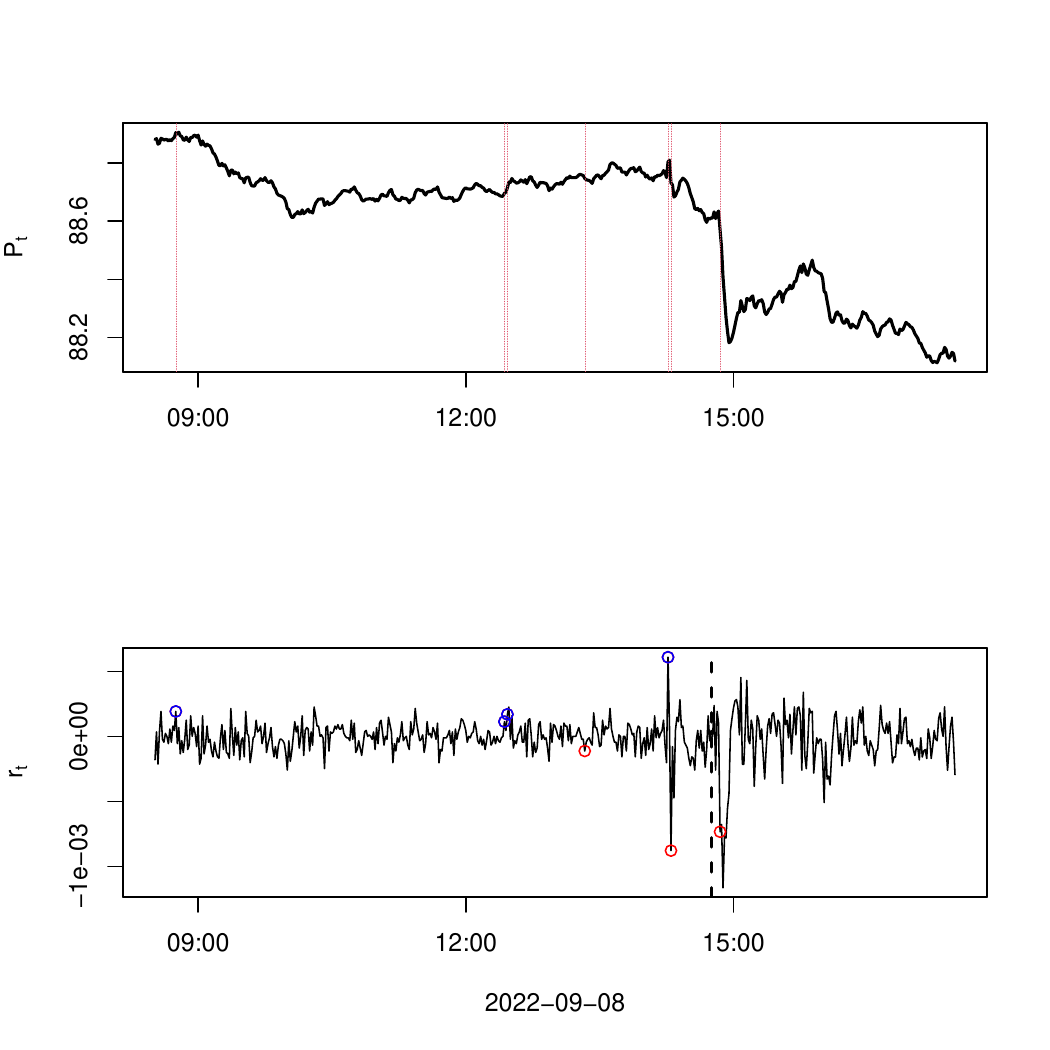}
	\end{subfigure}
	\hfill
	\begin{subfigure}[b]{0.5\textwidth}
		\includegraphics[width=\textwidth]{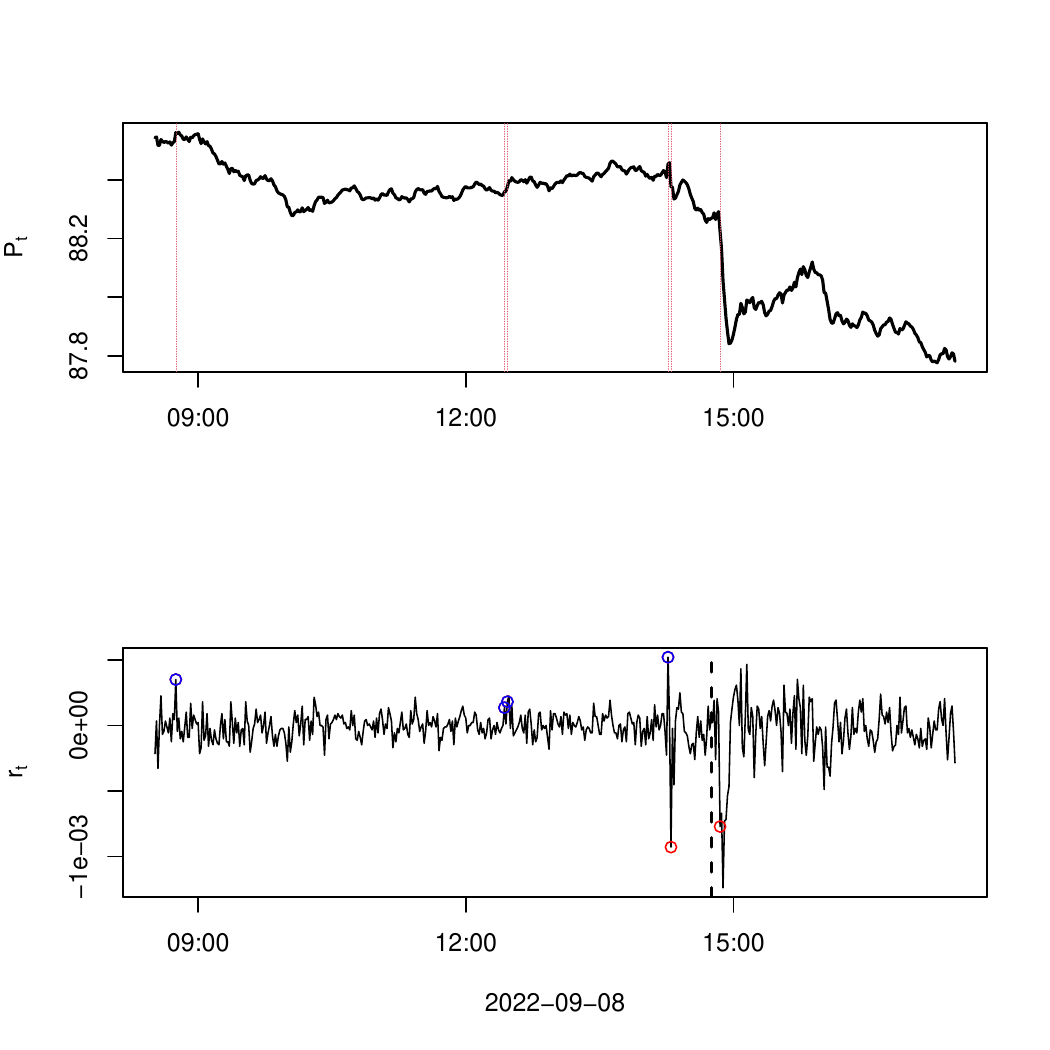}
	\end{subfigure}
	\caption{Prices and returns of the brown bullet bond issued by Engie SA  observed at September 8, 2022 (plots on the left refer to ask prices/returns while plots on the right to the bid counterparts). Red and blue circles refer respectively to negative and positive jumps in the series of returns identified through the LM test for $\alpha_{LM}=0.95$. Vertical red lines in the upper plots refer to time instants where (positive or negative) jumps occur. \label{fig:ds_ENGIFP_0_375_2027_corp_bro_OLD_RIGHT}}
\end{figure}

\begin{figure}[htbp]
	\begin{subfigure}[b]{0.5\textwidth}
		\centering
		\includegraphics[width=\textwidth]{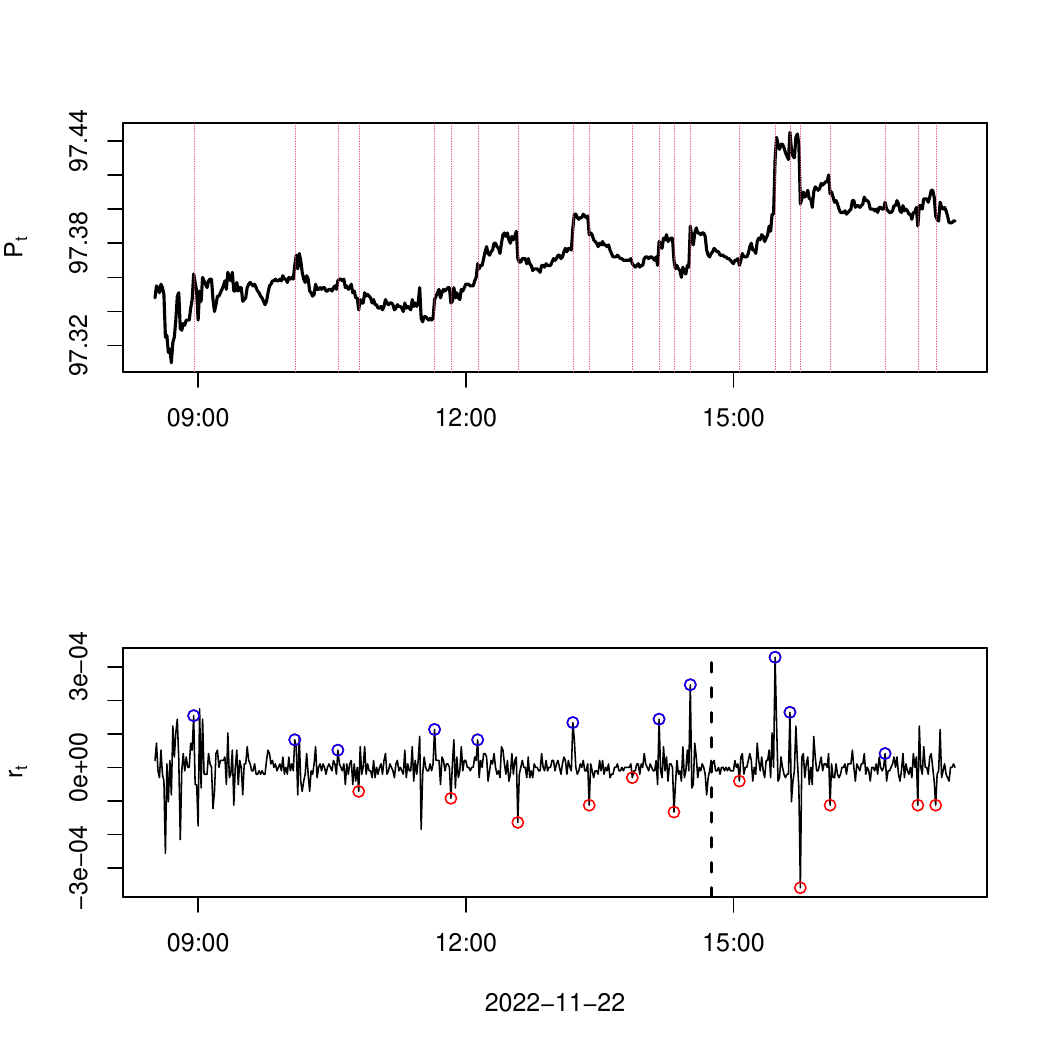}
	\end{subfigure}
	\hfill
	\begin{subfigure}[b]{0.5\textwidth}
		\includegraphics[width=\textwidth]{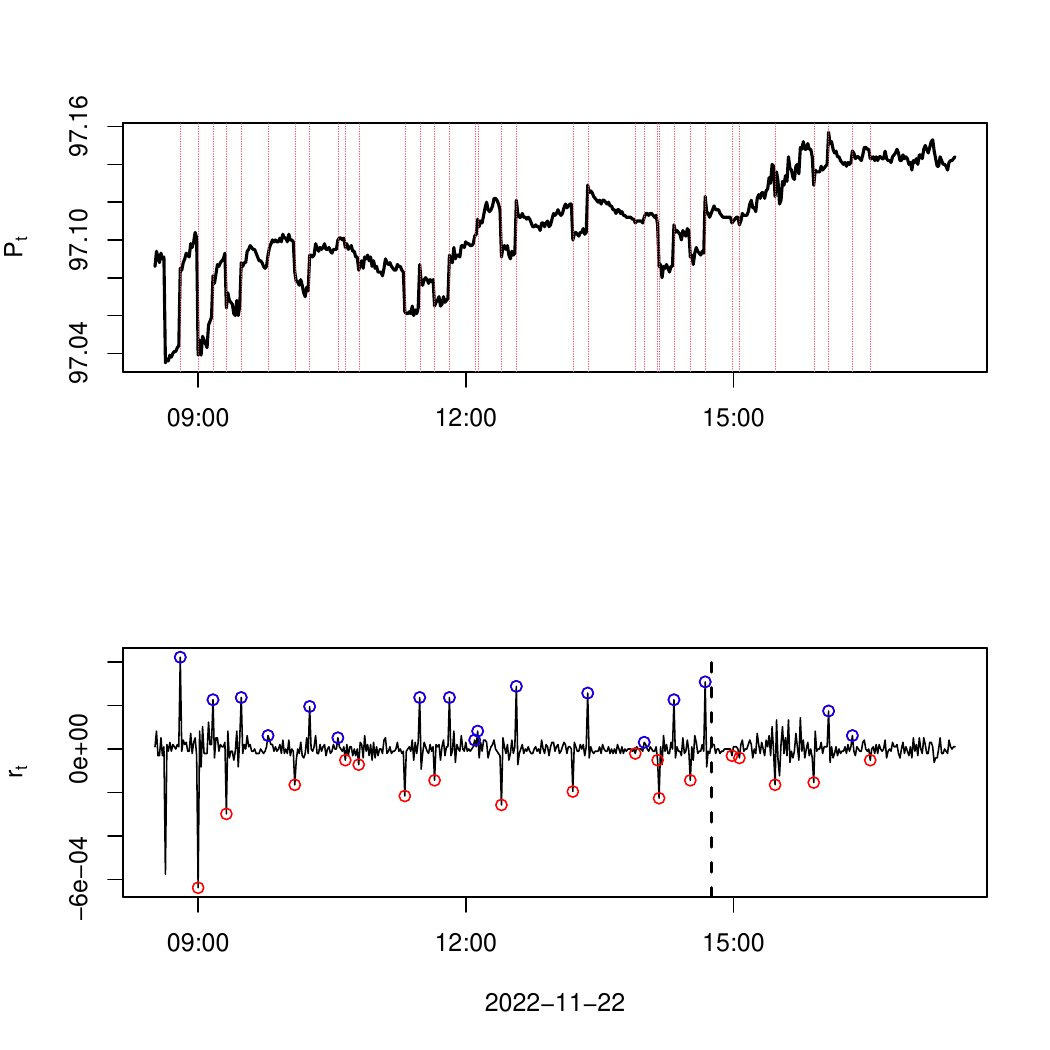}
	\end{subfigure}
	\caption{Prices and returns of the green bullet bond issued by Engie SA observed at November 22, 2022 (left ask; right bid). Red and blue circles refer respectively to negative and positive jumps in the series of returns identified through the LM test for $\alpha_{LM}=0.95$. Vertical red lines in the upper plots refer to time instants where (positive or negative) jumps occur. \label{fig:ds_ENGIFP_0_875_2024_corp_green_RECENT_RIGHT}}
\end{figure}

\begin{figure}[htbp]
	\begin{subfigure}[b]{0.5\textwidth}
		\centering
		\includegraphics[width=\textwidth]{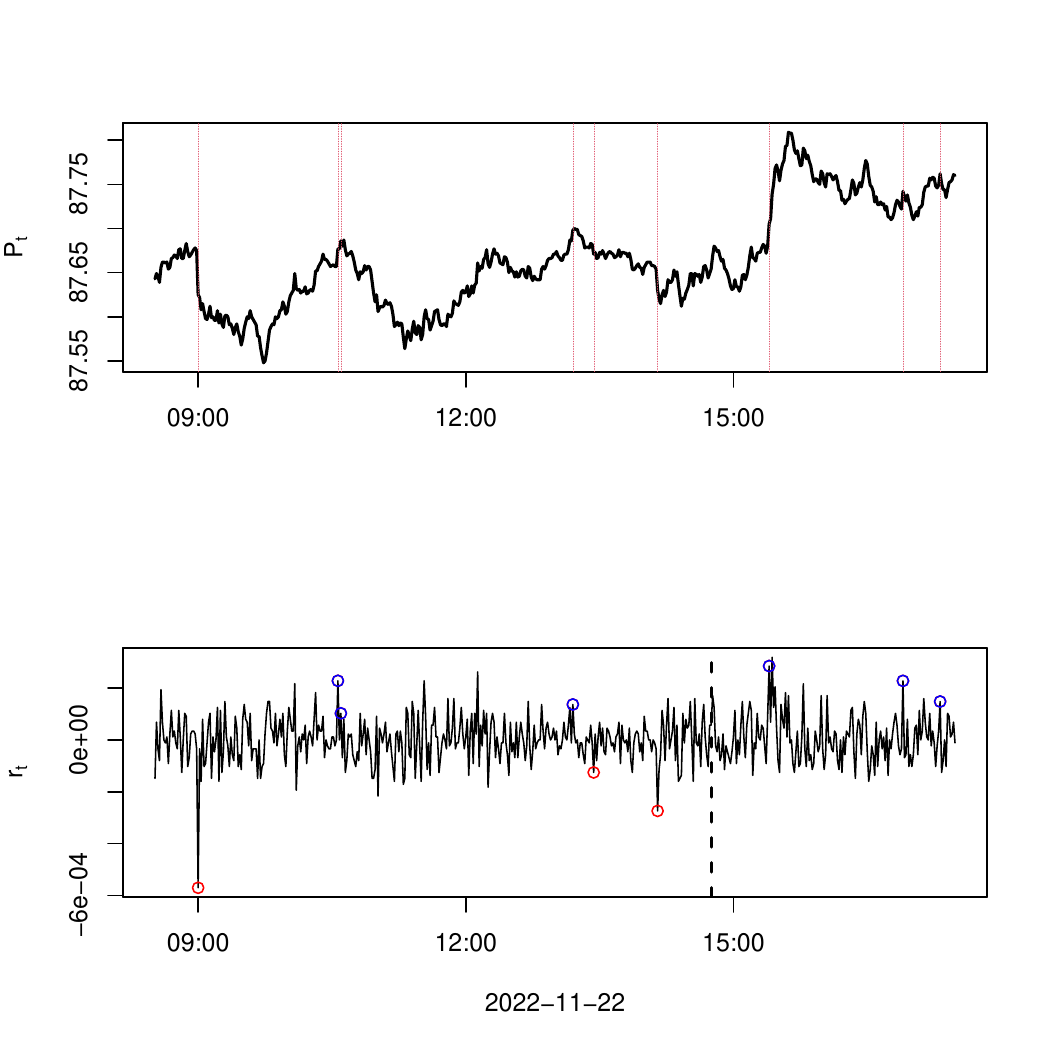}
	\end{subfigure}
	\hfill
	\begin{subfigure}[b]{0.5\textwidth}
		\includegraphics[width=\textwidth]{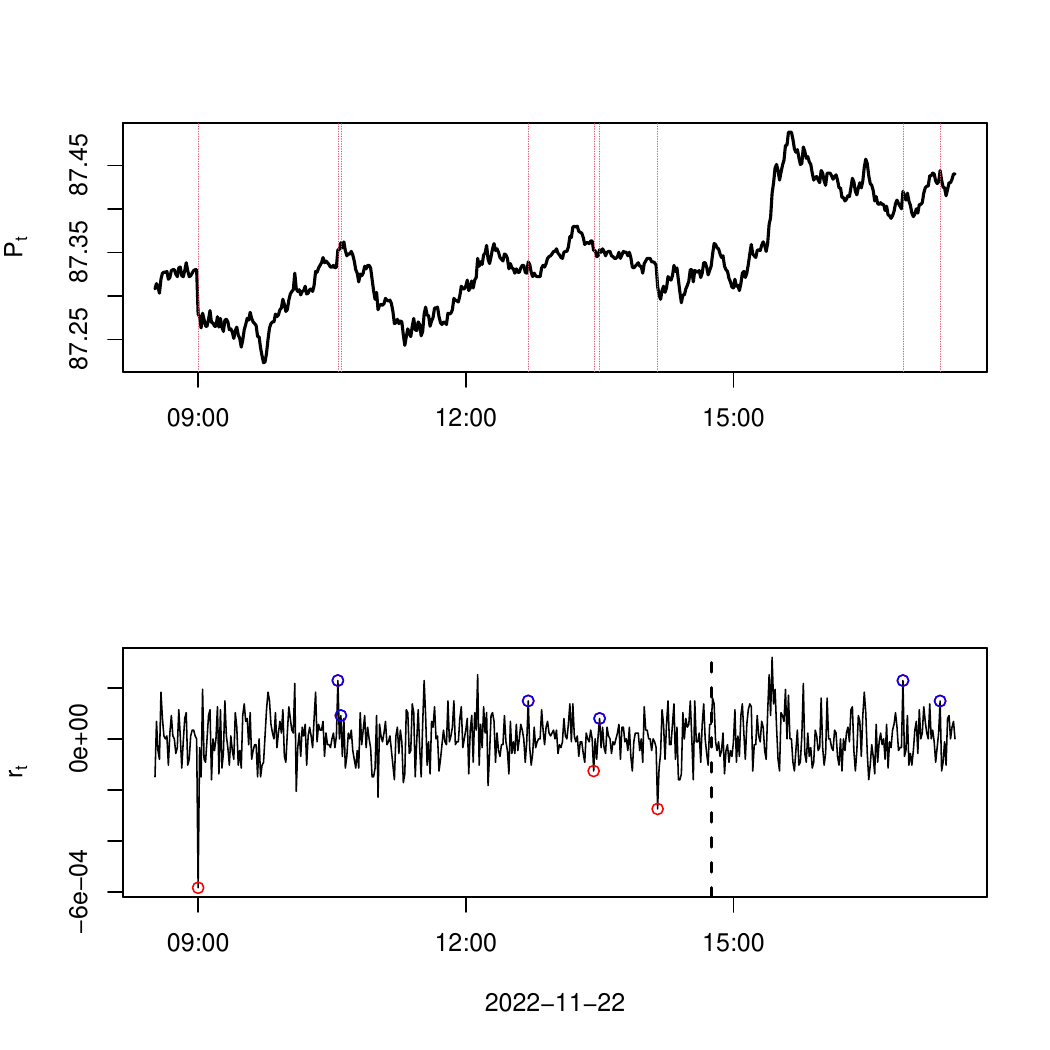}
	\end{subfigure}
	\caption{Prices and returns of the brown bullet bond issued by Engie SA observed at November 22, 2022 (left ask; right bid). Red and blue circles refer respectively to negative and positive jumps in the series of returns identified through the LM test for $\alpha_{LM}=0.95$. Vertical red lines in the upper plots refer to time instants where (positive or negative) jumps occur. \label{fig:ds_ENGIFP_0_375_2027_corp_bro_RECENT_RIGHT}}
\end{figure}

In this section we present the main results from the fitting of the univariate and bivariate CARMA(p,q)-Hawkes model introduced in Section \ref{modelsjd}. As the three frameworks under consideration require an underlying point process we consider the LM test for the jump identification. Indeed, for each series (bid and ask) we identify four sequences of jumps as we vary the confidence level of the LM test ($\alpha_{LM} \in \{0.95, 0.975, 0.99, 0.995\}$). For each value of $\alpha_{LM}$ we compute the percentage of observations classified as jumps. Data refer to intraday  bid and ask prices. \newline
Differences emerge both by looking at different bond types and different points in time. For example, Figures \ref{fig:ds_old_ACAFP_0_375_10_21_2025_GB} and \ref{fig:ds_old_ACAFP_0_5_2024_BB} refer to prices and returns of respectively the green and the brown bonds issued by Cr\'edit Agricole SA  (included in the dataset) observed at September 8, 2022. At this date, at 14:45 the ECB announced an increase in the reference interest rate. Both series (bid and ask) experienced a drop in the price value immediately after the ECB announcement. The same quantities for November 22, 2022 are displayed respectively in Figures \ref{fig:ds_rec_ACAFP_0_375_10_21_2025_GB} and \ref{fig:ds_rec_ACAFP_0_5_2024_BB}. Similar plots for callable bonds issued by Cr\'edit Agricole SA  and bullet bonds issued by Engie SA are reported in \ref{callableRes}. In particular Figures \ref{fig:ds_old_ACAFP_3_95_2032_GB_CA_todo_M} and \ref{fig:ds_old_ACAFP_3_8_2031_BB_CA} provide information on the set of callable bonds at September 8, 2022 while  \ref{fig:ds_rec_ACAFP_3_95_2032_GB_CA_todo_M} and \ref{fig:ds_rec_ACAFP_3_8_2031_BB_CA}  at November 22, 22. For the bonds issued by Engie SA we refer to Figures  \ref{fig:ds_ENGIFP_0_875_2024_corp_green_OLD_RIGHT} and \ref{fig:ds_ENGIFP_0_375_2027_corp_bro_OLD_RIGHT} for data observed at September 8, 20022 and to Figures \ref{fig:ds_ENGIFP_0_875_2024_corp_green_RECENT_RIGHT} and \ref{fig:ds_ENGIFP_0_375_2027_corp_bro_RECENT_RIGHT}  for November 22, 2022.

Our first attempt is devoted to the fitting of the bivariate CARMA(p,q)-Hawkes  model to the counting processes that are made up of any price down or up movement, i.e. framework 1.  The residual analysis suggests that the fitting made directly on these two series is not appropriate. We thus consider framework 2 that implies fitting a univariate CARMA(p,q)-Hawkes model to the counting processes obtained applying the LM test to the series of returns for varying levels of the confidence level $\alpha_{LM}$. Tables \ref{Res_Diff_CarmaHawkesProcessBid} and \ref{Res_Diff_CarmaHawkesProcessAsk} provide the $p$ and $q$ orders of the best fitting univariate CARMA(p,q)-Hawkes models respectively bid and ask prices for the green (bullet and callable) bonds issued by Cr\'edit Agricole SA . The last columns of these tables display the percentage of observations classified as jumps for varying levels of the confidence parameter in the LM test.  The results of the selection procedure  for the brown (bullet and callable) bonds issued by Cr\'edit Agricole SA  are reported respectively in Tables \ref{Res_Diff_CarmaHawkesProcessBROWNBid} and \ref{Res_Diff_CarmaHawkesProcessBROWNAsk} for bid and ask prices.  The corresponding results for green and brown bonds issued by Engie SA are presented respectively in Tables \ref{greencorp} and \ref{Browncorpbiv}.
During the trading period August 19-September 12, 2022  only around 4\% of observed returns are classified as jumps both for the brown and the green bullet bonds (bid series) while for the series based on ask prices in the case of brown bullet bonds we have 10\% of observations classified as jumps. In the second period under investigation, i.e. November 14-December 13, 2022 differences appear between the two categories of bullet bonds issued by Cr\'edit Agricole SA. Indeed, for the case of the green bond less than 5\% of observations are classified as jumps while for the brown bond this quantity is larger than 10\% especially in the case of the series based on bid prices. Series based on callable bond prices display totally different patterns as more than half of the observations are classified as jumps.

\begin{table}[!htbp]
\centering     
\footnotesize{\begin{tabular}{lccc}
\hline\hline
$\alpha_{LM}$      & \textbf{Orders p and q} & \textbf{KS Test ($p$-value)} & $\%$ \textbf{of Jumps} \\
\hline
 \multicolumn{4}{c}{\textsl{Bullet Green coup. 0.375$\%$ (Period I)}} \\
\hline
 95$\%$    &  (1,0) &  0.099 (0.074) &   4.59$\%$ \\
 97.5$\%$  &  (1,0) &  0.098 (0.067) &   4.83$\%$ \\
 99$\%$    &  (1,0) &  0.098 (0.068) &   4.83$\%$ \\
 99.5$\%$  &  (1,0) &  0.098 (0.049) &   5.24$\%$ \\
\hline
 \multicolumn{4}{c}{\textsl{Callable Green coup. 3.95$\%$ (Period I)}}\\
\hline
95$\%$    & (1,0)  & 0.079 (0.187) & 47.25$\%$\\
97.5$\%$  & (2,0)  & 0.042 (0.880) & 49.75$\%$\\
99$\%$    & (1,0)  & 0.093 (0.054) & 52.5$\%$\\
99.5$\%$  & (1,0)  & 0.097 (0.032) & 54.75$\%$\\
\hline
\multicolumn{4}{c}{\textsl{Bullet Green coup. 0.375$\%$ (Period II)}} \\
\hline
 95$\%$    & (1,0) & 0.073 (0.498)                  & 3.79$\%$ \\ 
 97.5$\%$  & (1,0) & 0.069 (0.534)                   & 3.93$\%$ \\
 99$\%$    & (1,0) & 0.064 (0.595)                  & 4.28$\%$ \\
 99$\%$    & (2,0) & 0.441 ($<$0.001)  & 4.34$\%$ \\
\hline 
\multicolumn{4}{c}{\textsl{Callable Green coup. 3.95$\%$ (Period II)}}\\
\hline
95$\%$   &  (2,0) & 0.106 (0.022)  &  70.88$\%$ \\
97.5$\%$ &  (2,0) & 0.132 (0.001)  &  76.14$\%$ \\
99$\%$   &  (2,0) & 0.123 (0.002)  &  81.05$\%$ \\
99$\%$   &  (2,0) & 0.136 ($<$0.001)   &  85.26$\%$ \\
\hline\hline
\end{tabular}}
\caption{Best fitting CARMA(p,q)-Hawkes  models for tick bid prices of green bonds (Bullet and Callable) issued by Cr\'edit Agricole SA  observed during two time intervals: August 19-September 19, 2022 (Period I) and November 14 - December 13, 2022 (Period II) for each $\alpha_{LM}$ in the LM test for the identification of jumps. The third and the fourth column refer respectively to the value of the KS test computed on the residuals of the best fitting model (in brackets its $p$-value) and the rate of observations classified as jumps.  \label{Res_Diff_CarmaHawkesProcessBid}}
\end{table}

\begin{table}[!htbp]
\centering     
\footnotesize{\begin{tabular}{lccc}
\hline\hline
$\alpha_{LM}$      & \textbf{Orders p and q} & \textbf{KS Test ($p$-value)} & $\%$ \textbf{of Jumps} \\
\hline
 \multicolumn{4}{c}{\textsl{Bullet Green coup. 0.375$\%$ (Period I)}} \\
\hline
  95$\%$    & (1,0) & 0.137 (0.008) & 3.87$\%$\\ 
  97.5$\%$  & (1,0) & 0.143 (0.003) & 4.22$\%$\\ 
  99$\%$    & (1,0) & 0.140 (0.003)  & 4.48$\%$\\ 
  99.5$\%$  & (1,0) & 0.139 (0.003) & 4.54$\%$\\ 
\hline
 \multicolumn{4}{c}{\textsl{Callable Green coup. 3.95$\%$ (Period I)}}\\
\hline
95$\%$    &  (1,0) & 0.098 (0.041) & 51.14$\%$\\ 
97.5$\%$  &  (1,0) & 0.093 (0.047) & 54.68$\%$\\
99$\%$    &  (2,0) & 0.077 (0.137) & 57.22$\%$\\
99.5$\%$  &  (2,1) & 0.048 (0.648) & 60.00$\%$\\
\hline
 \multicolumn{4}{c}{\textsl{Bullet Green coup. 0.375$\%$ (Period II)}} \\
\hline
  95$\%$    & (2,0) & 0.366 ($<$0.001)  & 2.64$\%$\\ 
  97.5$\%$  & (2,0) & 0.366 ($<$0.001)  & 2.64$\%$\\ 
  99$\%$    & (2,0) & 0.272 ($<$0.001) & 2.90$\%$\\ 
  99.5$\%$  & (2,0) & 0.281 ($<$0.001)  & 2.96$\%$\\ 
\hline 
\multicolumn{4}{c}{\textsl{Callable Green coup. 3.95$\%$ (Period II)}}\\
\hline
  95$\%$    & (2,0) & 0.072 (0.275) & 71.33$\%$\\ %
  97.5$\%$  & (2,0) & 0.063 (0.414)  & 76.00$\%$\\ 
  99$\%$    & (2,0) & 0.221 ($<$0.001)  & 76.41$\%$\\ 
  99.5$\%$  & (2,0) & 0.162 ($<$0.001)  & 80.28$\%$\\ 
\hline\hline
\end{tabular}}
\caption{Best fitting CARMA(p,q)-Hawkes models for tick ask prices of green bonds (bullet and callable) issued by Cr\'edit Agricole SA  observed during two time intervals: August 19-September 19, 2022 (Period I) and November 14 - December 13, 2022 (Period II) for each $\alpha_{LM}$ in the LM test for the identification of jumps. The third and the fourth column refer respectively to the value of the KS test computed on the residuals of the best fitting model (in brackets its $p$-value) and the rate of observations classified as jumps. \label{Res_Diff_CarmaHawkesProcessAsk}}
\end{table}

\begin{table}[!htbp]
\centering     
\footnotesize{\begin{tabular}{lccc}
\hline\hline
$\alpha_{LM}$      & \textbf{Orders p and q} & \textbf{KS Test ($p$-value)} & $\%$ \textbf{of Jumps} \\
\hline
 \multicolumn{4}{c}{\textsl{Bullet Brown coup. 0.5$\%$ (Period I)}} \\
\hline
 95$\%$    &    (1,0) & 0.064 (0.524) & 4.47$\%$\\ 
 97.5$\%$    &  (1,0) & 0.067 (0.418) & 4.83$\%$\\ 
 99$\%$    &    (1,0) & 0.064 (0.412) & 5.3$\%$\\ 
 99.5$\%$    &  (1,0) & 0.065 (0.357) & 5.54$\%$\\
\hline
 \multicolumn{4}{c}{\textsl{Callable Brown coup. 3.8$\%$ (Period I)}}\\
\hline
 95$\%$    &     (1,0) & 0.102 (0.091) & 42.74$\%$\\ 
 97.5$\%$    &  (2,0)  & 0.103 (0.072) & 45.30$\%$\\ 
 99$\%$    &  (1,0)    & 0.109 (0.040) & 47.29$\%$\\
 99.5$\%$    &  (2,0)  & 0.076 (0.257) & 50.43$\%$\\
\hline
\multicolumn{4}{c}{\textsl{Bullet Brown Bond coup. 0.5$\%$ (Period II)}} \\
\hline
 95$\%$    & (2,0)   & 0.043 (0.255) & 16.24$\%$\\
 97.5$\%$    & (2,0) & 0.039 (0.310) & 16.87$\%$\\
 99$\%$    & (2,0)   & 0.036 (0.394) & 17.42$\%$\\
 99.5$\%$    & (2,0) & 0.042 (0.225) & 17.94$\%$\\
\hline 
\multicolumn{4}{c}{Callable Brown coup. 3.8$\%$ (Period II) }\\
\hline
 95$\%$ & (2,0)   & 0.185 ($<$0.001) & 72.08$\%$\\
 97.5$\%$ & (2,0) & 0.192 ($<$0.001) & 74.11$\%$\\
 99$\%$ & (2,0)   & 0.208 ($<$0.001) & 78.17$\%$\\
 99.5$\%$ & (2,0) & 0.194 ($<$0.001) & 79.7$\%$\\
\hline\hline
\end{tabular}}
\caption{Best fitting CARMA(p,q)-Hawkes models for tick bid prices of brown bonds (bullet and callable) issued by Cr\'edit Agricole SA  observed during two time intervals: August 19-September 19, 2022 (Period I) and November 14 - December 13, 2022 (Period II) for each $\alpha_{LM}$ in the LM test for the identification of jumps. The third and the fourth column refer respectively to the value of the KS test computed on the residuals of the best fitting model (in brackets its $p$-value) and the rate of observations classified as jumps. \label{Res_Diff_CarmaHawkesProcessBROWNBid}}
\end{table}

\label{Res_Diff_CarmaHawkesProcessBROWNBid}

\begin{table}[!htbp]
\centering     
\footnotesize{\begin{tabular}{lccc}
\hline\hline
$\alpha_{LM}$      & \textbf{Orders p and q} & \textbf{KS Test ($p$-value)} & $\%$ \textbf{of Jumps} \\
\hline
 \multicolumn{4}{c}{\textsl{Bullet Brown coup. 0.5$\%$ (Period I) }} \\
\hline
 95$\%$    & (1,0) &  0.069 (0.070) & 9.88$\%$ \\ 
 97.5$\%$  & (1,0) &  0.068 (0.062) & 10.32$\%$ \\ 
 99$\%$ & (1,0)   &  0.071 (0.039) & 10.79$\%$ \\
 99.5$\%$ & (1,0) &  0.071 (0.038) & 11.01$\%$ \\
\hline
 \multicolumn{4}{c}{\textsl{Callable Brown coup. 3.8$\%$ (Period I) }}\\
\hline
95$\%$   & (2,0) &  0.037 (0.965) & 50.28$\%$ \\%
97.5$\%$ & (2,1) &  0.032 (0.992) & 51.69$\%$ \\%
99$\%$   & (1,0) &  0.079 (0.180) & 54.24$\%$ \\%
99.5$\%$ & (1,0) &  0.084 (0.129) & 55.65$\%$ \\%
\hline
\multicolumn{4}{c}{\textsl{Bullet Brown coup. 0.5$\%$ (Period II)} } \\
\hline
 95$\%$   & (1,0) &  0.046  (0.388)  & 11.63$\%$ \\
 97.5$\%$ & (1,0) &  0.050  (0.282) & 11.96$\%$ \\
 99$\%$   & (2,0) &  0.031  (0.812) & 12.32$\%$ \\
 99.5$\%$ & (2,0) &  0.032  (0.783)  & 12.44$\%$ \\
\hline 
\multicolumn{4}{c}{\textsl{Callable Brown  coup. 3.8$\%$ (Period II)}}\\
\hline
95$\%$   & (2,0) & 0.126 (0.022) & 66.82$\%$ \\%
97.5$\%$ & (2,0) & 0.147 (0.003) & 70.09$\%$ \\
99$\%$   & (2,1) & 0.123 (0.020) & 71.96$\%$ \\%
99.5$\%$ & (2,1) & 0.054 (0.746) & 73.83$\%$ \\%
\hline\hline
\end{tabular}}
\caption{Best fitting CARMA(p,q)-Hawkes models for a dataset based on tick ask prices of brown bonds (bullet and callable) issued by Cr\'edit Agricole SA  observed during two time intervals: August 19-September 19, 2022 (Period I) and November 14 - December 13, 2022 (Period II) for each $\alpha_{LM}$ in the LM test for the identification of jumps. The third and the fourth column refer respectively to the value of the KS test computed on the residuals of the best fitting model (in brackets its $p$-value) and the rate of observations classified as jumps.\label{Res_Diff_CarmaHawkesProcessBROWNAsk}}
\end{table}

For the cases where the KS fails, we test framework 3 i.e. we split jumps identified through the LM test into positive and negative ones. We then apply the bivariate CARMA(p,q)-Hawkes model to the two counting processes. The results are reported in Tables  \ref{LastModelForAllCasesGB} and \ref{LastModelForAllCasesBB} for green and brown bonds respectively. We observe that in many cases even this level of deepness of the analysis is not sufficient to produce a statistically significant model that can reproduce our dataset. The opposite happens in the case of the green and brown bonds issued by Engie SA as we are able to find in most of the cases a model that, based on the residual analysis, seems adequate for fit our data (see Tables \ref{greencorp3} and \ref{browncorp3}).

\begin{table}[!htbp]
\centering     
\footnotesize{\begin{tabular}{lccc}
\hline\hline
$\alpha_{LM}$   & \textbf{Orders p and q}  & \textbf{KS$_1$ ($p$-value)} & \textbf{KS$_2$ ($p$-value)}\\
\hline 
\multicolumn{4}{c}{\textsl{Bullet Green coup. 0.375$\%$ (Period I) Bid}} \\
\hline
99.5$\%$  &  $\left(1,1,0,0,0,0\right)$ & 0.142 (0.154) & 0.065 (0.859)\\ 
\hline
\multicolumn{4}{c}{\textsl{Callable Green coup. 3.95$\%$ (Period I) Bid}} \\
\hline
99.5$\%$  & $\left(2,2,1,0,1,0\right)$ & 0.074 (0.546) & 0.065 (0.735)\\
\hline
\multicolumn{4}{c}{\textsl{Callable Green coup. 3.95$\%$ (Period II) Bid}} \\
\hline
95$\%$    & $\left(2,2,1,0,0,0\right)$ & 0.116 (0.056) & 0.354 (2.34e-08)\\
97.5$\%$  & $\left(2,2,1,0,0,0\right)$ & 0.138 (0.008) & 0.129 (0.145)\\
99$\%$    & $\left(2,2,1,0,0,0\right)$ & 0.097 (0.117) & 0.116 (0.196)\\
99.5$\%$  & $\left(2,1,1,0,1,0\right)$ & 0.119 (0.024) & 0.093 (0.406)\\
\hline
\multicolumn{4}{c}{\textsl{Callable Green coup. 3.95$\%$  (Period I) Ask}}\\
\hline
97.5$\%$  &  $\left(2,2,0,0,0,0\right)$ & 0.061 (0.825) & 0.079 (0.519)\\
\hline
\multicolumn{4}{c}{\textsl{Bullet Green coup. 0.375$\%$(Period II) Ask}} \\
\hline
97.5$\%$   & $\left(2,2,1,1,1,1\right)$ & 0.177 (0.011) &  0.109 (0.32) \\
99$\%$     & $\left(2,1,1,1,0,0\right)$ & 0.171 (0.0132) &  0.135 (0.097) \\
99.5$\%$   & $\left(2,2,1,1,1,1\right)$ & 0.194 (0.003) &  0.099 (0.369) \\
\hline
\multicolumn{4}{c}{\textsl{Bullet Green coup. 0.375$\%$(Period II)  Ask}} \\
\hline
95$\%$    &  $\left(1,1,0,0,0,0\right)$  & 0.103 (0.691) & 0.162 (0.196)\\ 
97.5$\%$  &  $\left(1,1,0,0,0,0\right)$  & 0.103 (0.691) & 0.162 (0.196)\\ 
99$\%$    &  $\left(1,1,0,0,0,0\right)$  & 0.154 (0.159) & 0.143 (0.285)\\
99.5$\%$  &  $\left(1,1,0,0,0,0\right)$  & 0.179 (0.064) & 0.121 (0.459)\\
\hline
\multicolumn{4}{c}{\textsl{Callable Green coup. 3.95$\%$  (Period II) Ask}}\\
\hline
99$\%$   &  $\left(2,2,1,0,0,1\right)$  & 0.115 (0.060) & 0.148 (0.062) \\
99.5$\%$ &  $\left(2,2,1,0,0,1\right)$  & 0.099 (0.126) & 0.119 (0.178) \\
\hline\hline
\end{tabular}}
\caption{Results for a a bivariate CARMA(p,q)-Hawkes model as specified in framework 3 presented in Section \ref{modelsjd} fitted to a set tick ask prices of green bonds (bullet and callable) issued by Cr\'edit Agricole SA  observed during two time intervals: August 19-September 19, 2022 (Period I) and November 14 - December 13, 2022 (Period II) for the levels of $\alpha_{LM}$ in the LM test. 
The third and the fourth column refer respectively to the value of the KS test computed on the residuals (in brackets its $p$-value) from the set of upward and downward jumps.\label{LastModelForAllCasesGB}
}
\end{table}

\begin{table}[!htbp]
\centering     
\footnotesize{\begin{tabular}{lccc}
\hline\hline
$\alpha_{LM}$ &  \textbf{Orders p and q} & \textbf{KS$_1$ ($p$-value)} & \textbf{KS$_2$ ($p$-value)}\\
\hline 
\multicolumn{4}{c}{\textsl{Callable Brown Bond - (Period I) - coupon rate 3.8$\%$ Bid}}\\
\hline
99$\%$ & $\left(2,2,1,0,0,1\right)$ & 0.067 (0.764) & 0.099 (0.381)\\
\hline 
\multicolumn{4}{c}{\textsl{Callable Brown Bond - (Period II) - coupon rate 3.8$\%$ Bid}}\\
\hline
95$\%$   & $\left(2, 2, 1,0, 0, 0\right)$ & 0.116 (0.226) & 0.146 (0.178) \\
97.5$\%$   & $\left(2,2,0,0, 0,0\right)$  & 0.089 (0.510) & 0.155 (0.110) \\
99$\%$   & $\left(2,2,0,0,0,0\right)$     & 0.090 (0.462) & 0.153 (0.109) \\
99.5$\%$   & $\left(2,2,1,0,0,0\right)$   & 0.114 (0.194) & 0.154 (0.091) \\
\hline
\multicolumn{4}{c}{\textsl{Bullet Brown Bond - (Period I) - coupon rate 0.5$\%$}} \\
\hline
99$\%$ & $\left(2,1,1,0,1,0\right)$  &  0.219 ($<$0.001) &  0.070  (0.287) \\ 
99.5$\%$ & $\left(2,1,1,0,1,0\right)$ &  0.276 ($<$0.001) &  0.1697  ($<$0.001) \\ 
\hline
 \multicolumn{4}{c}{\textsl{Callable Brown Bond - (Period II) - coupon rate 3.8$\%$ Ask}}\\
\hline
95$\%$ & $\left(2,2,1,0,0,0\right)$   &  0.064 (0.872) & 0.155 (0.157) \\
97.5$\%$ & $\left(2,2,0,0,0,0\right)$ &  0.072 (0.728) & 0.149 (0.163) \\
99$\%$ & $\left(2,2,0,0,0,0\right)$   &  0.132 (0.077) & 0.170 (0.070) \\
\hline\hline
\end{tabular}}
\caption{Results for a bivariate CARMA(p,q)-Hawkes model as specified in framework 3 presented in Section \ref{modelsjd} fitted to a set tick ask prices of brown bonds (bullet and callable) issued by Cr\'edit Agricole SA  observed during two time intervals: August 19-September 19, 2022 (Period I) and November 14 - December 13, 2022 (Period II) for the levels of $\alpha_{LM}$ in the LM test. 
The third and the fourth column refer respectively to the value of the KS Test computed on the residuals (in brackets its $p$-value) from the set of upward and downward jumps. \label{LastModelForAllCasesBB}
}
\end{table}

\color{black}
\clearpage

\section{Conclusions}
\label{concl}
In this paper we investigated market activity of bonds through the introduction of a bivariate model for the intensity of positive and negative jumps in a high frequency framework where we allow for a non-monotonic vanishing effect of jumps. We considered both ask and bid prices where we identified the best fitting models for brown and green bonds. This study dealt both with bullet and callable bonds as we expected these two instruments to behave differently when observed at the lens of trading activity.
Results suggest that the bid-ask spread of green bonds is larger on average, especially during autumn. Moreover, higher order CARMA(p,q)-Hawkes models seem to be useful in order to reproduce the sequence of jumps.

\bibliography{mybibfile}
\appendix

\section{Likelihood function}
\label{appendice}
\noindent The solution of $\mathrm{d} \mathbf{X}_t  = \mathbf{A} \mathbf{X}_t \mathrm{d}t + \mathbf{e}\mathrm{d}\mathbf{N}_t, \; \text{with} \; \mathbf{X}_0 = \mathbf{0}$ (linear SDE) is given as in \eqref{solXt} and substituting it in the first integral of \eqref{likelihood}, the likelihood function can be rewritten as
\begin{eqnarray*}
\mathcal{L} &=&  - \mathbbm{1}^\intercal \bm{\mu} \left(T - 0 \right) - \mathbbm{1}^\intercal \mathbf{B} \int_{0}^{T} \left[ \int_{0}^{t} e^{\mathbf{A} \left(t - s \right)} \mathbf{e}\mathrm{d}\mathbf{N}_s  \right]\mathrm{d}t + \displaystyle \sum_{T_k \leq T}   \ln \left(\bm{\mu} + \mathbf{B} \mathbf{X}_{T_k} \right)^\intercal \Delta \mathbf{N}_{T_k} \\ 
 &= & - \mathbbm{1}^\intercal \bm{\mu}T  - \mathbbm{1}^\intercal \mathbf{B} \int_{0}^{T} \mathbf{A}^{-1} \left[e^{\mathbf{A} \left(T - s \right)}   - \mathbf{I} \right] \mathbf{e}\mathrm{d}\mathbf{N}_s + \displaystyle \sum_{T_k \leq T}   \ln \left(\bm{\mu} + \mathbf{B} \mathbf{X}_{T_k} \right)^\intercal \Delta \mathbf{N}_{T_k}\\
& = &  - \mathbbm{1}^\intercal \bm{\mu}T  - \mathbbm{1}^\intercal \mathbf{B}\mathbf{A}^{-1} \displaystyle \sum_{T_k \leq T}  e^{\mathbf{A} \left(T - T_k \right)}\mathbf{e}\Delta \mathbf{N}_{T_k}   +  \mathbbm{1}^\intercal \mathbf{B}\mathbf{A}^{-1} \mathbf{e} \displaystyle \sum_{T_k 	\leq T} \Delta \mathbf{N}_{T_k}\\
& +& \displaystyle \sum_{T_k \leq T}   \ln \left(\bm{\mu}
+\mathbf{B} \mathbf{X}_{T_k} \right)^\intercal \Delta \mathbf{N}_{T_k}\\
\end{eqnarray*}
where $\mathbbm{1}^\intercal \mathbf{B}\mathbf{A}^{-1} \mathbf{e} \displaystyle \sum_{T_k \leq T} \Delta \mathbf{N}_{T_k} = \mathbbm{1}^\intercal \mathbf{B}\mathbf{A}^{-1} \mathbf{e} \mathbf{N}_{T}$.
Therefore the likelihood function becomes
\begin{equation}\label{eq: liklihood}
\mathcal{L} =  - \mathbbm{1}^\intercal \bm{\mu}T + \mathbbm{1}^\intercal \mathbf{B}\mathbf{A}^{-1} \mathbf{e} \mathbf{N}_{T} - \mathbbm{1}^\intercal \mathbf{B}\mathbf{A}^{-1} \displaystyle \sum_{T_k \leq T}  e^{\mathbf{A} \left(T - T_k \right)}\mathbf{e}\Delta \mathbf{N}_{T_k}  + \displaystyle \sum_{T_k 	\leq T}   \ln \left(\bm{\mu} + \mathbf{B} \mathbf{X}_{T_k} \right)^\intercal \Delta \mathbf{N}_{T_k}.
\end{equation}
If we observe $n$ jumps in $\left[0, T\right]$, $\sum_{T_k \leq T}  e^{\mathbf{A} \left(T - T_k \right)} \mathbf{e}\Delta \mathbf{N}_{T_k} = e^{\mathbf{A} \left(T - T_n \right)} \displaystyle \sum_{k = 1}^{n} e^{\mathbf{A} \left(T_n - T_k \right)}\mathbf{e}\Delta \mathbf{N}_{T_k}$ and define $s(n) = \displaystyle \sum_{k = 1}^{n} e^{\mathbf{A} \left(T_n - T_k \right)} \mathbf{e}\Delta \mathbf{N}_{T_k}$, \eqref{eq: liklihood} reads
\begin{equation}
	\mathcal{L} =  - \mathbbm{1}^\intercal \bm{\mu}T + \mathbbm{1}^\intercal \mathbf{B}\mathbf{A}^{-1} \mathbf{e} \mathbf{N}_{T} - \mathbbm{1}^\intercal \mathbf{B}\mathbf{A}^{-1}  e^{\mathbf{A} \left(T - T_n \right)} s(n)  + \displaystyle \sum_{i = 1}^{n}   \ln \left(\bm{\mu} + \mathbf{B} \mathbf{X}_{T_i} \right)^\intercal \Delta \mathbf{N}_{T_i}.
\end{equation}
The final expression of the likelihood function is explicited assuming that $T = T_n$ i.e. we have $n$ jumps in $[0, T_n]$.
\section{Estimation procedure}
\clearpage
\section{Results for callable and corporate bonds}
\label{callableRes}

\clearpage
\begin{table}[!htbp]
	\centering     
	\footnotesize{\begin{tabular}{lccc}
			\hline\hline
			$\alpha_{LM}$      & \textbf{Orders p and q} & \textbf{KS Test ($p$-value)} & $\%$ \textbf{of Jumps} \\
			\hline
			\multicolumn{4}{c}{\textsl{Bullet Green coup. 0.875$\%$ (Period I)} Ask} \\
			\hline
			95$\%$    & (2,0) &  0.129 (0.010) & 35.29$\%$ \\ 
			97.5$\%$  & (2,1) &  0.111 (0.028) & 38.34$\%$ \\ 
			99$\%$    & (2,1) &  0.069 (0.309)& 42.48$\%$ \\
			99.5$\%$  & (2,1) &  0.056 (0.547) & 44.01$\%$ \\
			\hline
			\multicolumn{4}{c}{\textsl{Bullet Green coup. 0.875$\%$ (Period I) } Bid}\\
			\hline
			95$\%$   & (2,0) &  0.053 (0.474) & 58.64$\%$ \\%
			97.5$\%$ & (2,1) &  0.076 (0.085) & 63.18$\%$ \\%
			99$\%$   & (2,0) &  0.057 (0.298) & 65.68$\%$ \\%
			99.5$\%$ & (2,0) &  0.049 (0.472) & 68.18$\%$ \\%
			\hline
			\multicolumn{4}{c}{\textsl{Bullet Green coup. 0.875$\%$ (Period II)} Ask} \\
			\hline
			95$\%$   & (1,0) &  0.073  (0.165)   & 54.80$\%$ \\
			97.5$\%$ & (2,1) &  0.050  (0.567) & 57.85$\%$ \\
			99$\%$   & (2,0) &  0.035  (0.909) & 62.30$\%$ \\
			99.5$\%$ & (2,0) &  0.032  (0.941)    & 64.40$\%$ \\
			\hline 
			\multicolumn{4}{c}{\textsl{Bullet Green coup. 0.875$\%$ (Period II)} Bid}\\
			\hline
			95$\%$   & (2,1) & 0.046 (0.521) & 76.47$\%$ \\%
			97.5$\%$ & (2,1) & 0.044 (0.559) & 77.94$\%$ \\
			99$\%$   & (2,1) & 0.049 (0.415) & 81.13$\%$ \\%
			99.5$\%$ & (2,1) & 0.053 (0.309) & 82.60$\%$ \\%
			\hline\hline
	\end{tabular}}
	\caption{Best fitting CARMA(p,q)-Hawkes models for tick ask and bid prices of green bullet bonds issued by Engie SA observed during two time intervals: August 19-September 19, 2022 (Period I) and November 14 - December 13, 2022 (Period II) for each $\alpha_{LM}$ in the LM test for the identification of jumps. The third and the fourth column refer respectively to the value of the KS test computed on the residuals of the best fitting model (in brackets its $p$-value) and the rate of observations classified as jumps.\label{greencorp}}
\end{table}

\begin{table}[!htbp]
	\centering     
	\footnotesize{\begin{tabular}{lccc}
			\hline\hline
			$\alpha_{LM}$ &  \textbf{Orders p and q}  & \textbf{KS$_1$ ($p$-value)} & \textbf{KS$_2$ ($p$-value)}\\
			\hline 
			\multicolumn{4}{c}{\textsl{Bullet Green coup. 0.875$\%$ (Period I)} Ask}\\
			\hline
			95$\%$ & $\left(2,2,0,0,0,0\right)$   & 0.059 (0.911) & 0.070 ( 0.838)\\
			97.5$\%$ & $\left(2,2,0,0,0,0\right)$ & 0.077 (0.643) & 0.079 (0.638)\\
			\hline\hline
	\end{tabular}}
	\caption{Results for a bivariate CARMA(p,q)-Hawkes model as specified in framework 3 presented in Section \ref{modelsjd} fitted to a set of tick  bid and ask prices of green bullet bonds issued by Engie SA observed during two time intervals: August 19-September 19, 2022 (Period I) and November 14 - December 13, 2022 (Period II) for the levels of $\alpha_{LM}$ in the LM test. 
The third and the fourth column refer respectively to the value of the KS test computed on the residuals (in brackets its $p$-value) from the set of upward and downward jumps. \label{greencorp3}
	}
\end{table}

\begin{table}[!htbp]
	\centering     
	\footnotesize{\begin{tabular}{lccc}
			\hline\hline
			$\alpha_{LM}$      & \textbf{Orders p and q} & \textbf{KS Test ($p$-value)} & $\%$ \textbf{of Jumps} \\
			\hline
			\multicolumn{4}{c}{\textsl{Bullet Brown coup. 0.375$\%$ (Period I)} Ask} \\
			\hline
			95$\%$    & (1,0) &  0.064 (0.861) & 16.73$\%$ \\ 
			97.5$\%$  & (1,0) &  0.057 (0.911) & 18.53$\%$ \\ 
			99$\%$    & (1,0) &  0.067 (0.757)& 20.12$\%$ \\
			99.5$\%$  & (2,0) &  0.387 ($<$0.001) & 22.31$\%$ \\
			\hline
			\multicolumn{4}{c}{\textsl{Bullet Brown coup. 0.375$\%$ (Period I) } Bid}\\
			\hline
			95$\%$   & (2,0) &  0.365 ($<$0.001) & 20.28$\%$ \\%
			97.5$\%$ & (2,0) &  0.387 ($<$0.001) & 22.24$\%$ \\%
			99$\%$   & (1,0) &  0.074 (0.518) & 24.21$\%$ \\%
			99.5$\%$ & (1,0) &  0.089 (0.250) & 25.98$\%$ \\%
			\hline
			\multicolumn{4}{c}{\textsl{Bullet Brown coup. 0.375$\%$ (Period II)} Ask} \\
			\hline
			95$\%$   & (1,0) &  0.095  (0.2888)  & 21.34$\%$ \\
			97.5$\%$ & (2,0) &  0.432  ($<$0.001) & 23.72$\%$ \\
			99$\%$   & (2,0) &  0.438  ($<$0.001) & 27.27$\%$ \\
			99.5$\%$ & (1,0) &  0.091 (0.1701)  & 29.84$\%$ \\
			\hline 
			\multicolumn{4}{c}{\textsl{Bullet Brown coup. 0.375$\%$ (Period II)} Bid}\\
			\hline
			95$\%$   & (1,0) & 0.088 (0.330) & 22.90$\%$ \\%
			97.5$\%$ & (2,0) & 0.403 ($<$0.001) & 25.24$\%$ \\
			99$\%$   & (2,0) & 0.481 ($<$0.001) & 27.40$\%$ \\%
			99.5$\%$ & (1,0) & 0.069 (0.466) & 29.55$\%$ \\%
			\hline\hline
	\end{tabular}}
	\caption{Best fitting CARMA(p,q)-Hawkes models for tick ask and bid prices of brown bullet bonds issued by Engie SA observed during two time intervals: August 19-September 19, 2022 (Period I) and November 14 - December 13, 2022 (Period II) for each $\alpha_{LM}$ in the LM test for the identification of jumps. The third and the fourth column refer respectively to the value of the KS test computed on the residuals of the best fitting model (in brackets its $p$-value) and the rate of observations classified as jumps. \label{Browncorpbiv}}
\end{table}

\begin{table}[!htbp]
	\centering     
	\footnotesize{\begin{tabular}{lccc}
			\hline\hline
			$\alpha_{LM}$ &  \textbf{Orders p and q}  & \textbf{KS$_1$ ($p$-value)} & \textbf{KS$_2$ ($p$-value)}\\
			\hline 
			\multicolumn{4}{c}{\textsl{Bullet Brown coup. 0.375$\%$ (Period I)} Ask}\\
			\hline
			99.5$\%$ & $\left(2,1,0,0,0,0\right)$ & 0.111 (0.464) & 0.246 (0.003)\\
			\hline 
			\multicolumn{4}{c}{\textsl{Bullet Brown coup. 0.375$\%$ (Period I) } Bid}\\
			\hline
			95$\%$   & $\left(2, 1, 0,0, 0, 0\right)$ & 0.117 (0.406) & 0.118 (0.550) \\
			97.5$\%$   & $\left(2,1,1,0,0,0\right)$  & 0.072 (0.891) & 0.101 (0.677) \\
			\hline
			\multicolumn{4}{c}{\textsl{Bullet Brown coup. 0.375$\%$ (Period II)} Ask}  \\
			\hline
			97.5$\%$ & $\left(2,1,1,0,1,0\right)$  &  0.143 (0.127) &  0.128  (0.344) \\ 
			99$\%$ & $\left(2,1,1,0,1,0\right)$ &  0.122 (0.182) &  0.232  (0.004) \\ 
			\hline
			\multicolumn{4}{c}{\textsl{Bullet Brown coup. 0.375$\%$ (Period II)} Bid}\\
			\hline
			97.5$\%$ & $\left(2,1,1,0,0,0\right)$ &  0.140 (0.113) & 0.066 (0.960) \\
			99$\%$ & $\left(2,1,1,0,0,0\right)$   &  0.043 (0.997) & 0.160 (0.103) \\
			\hline\hline
	\end{tabular}}
	\caption{Results for a bivariate CARMA(p,q)-Hawkes model as specified in framework 3 presented in Section \ref{modelsjd} fitted to a set of tick ask and bid prices of brown bullet bonds issued by Engie SA observed during two time intervals: August 19-September 19, 2022 (Period I) and November 14 - December 13, 2022 (Period II) for the levels of $\alpha_{LM}$ in the LM test. 
The third and the fourth column refer respectively to the value of the KS test computed on the residuals (in brackets its $p$-value) from the set of upward and downward jumps. \label{browncorp3}
	}
\end{table}

\end{document}